\documentclass{article}
\usepackage{amsmath, amsfonts, mathtools, amsthm, geometry, subfiles, accents, hyperref, bookmark, todonotes, mdframed, graphicx, tcolorbox, subcaption, float, tikz-cd}
\usepackage{amssymb}
\usepackage[utf8]{inputenc}
\usepackage{xcolor}
\usepackage[normalem]{ulem}
\newif\ifshowchanges \showchangestrue
\ifshowchanges
  
  \newcommand{\dlt}[1]{{\color{red}\sout{#1}}}
\else
  
  \newcommand{\dlt}[1]{}
\fi
\usepackage{cleveref}
\usepackage[backend=biber,style=numeric,sorting=none]{biblatex}
\newcommand{\pleb}{Pleba\'nski}
\newcommand{\sd}{\Sigma}
\newcommand{\asd}{\bar{\Sigma}}

\newcommand{\C}{\mathbb{C}}
\newcommand{\R}{\mathbb{R}}

\newcommand{\im}{{\rm i}}

\newcommand{\so}{\mathfrak{so}}

\newcommand{\Etwoforms}{E \otimes{} \Lambda^2}
\newcommand{\Eoneforms}{E \otimes{} \Lambda^1}

\title{First-Order Structure of the Linearised Einstein Equations: \\ Constant-Curvature Backgrounds}
\author{Kirill Krasnov and Adam Shaw\\ 
School of Mathematical Sciences, University of Nottingham, Nottingham, NG7 2RD, UK}

\begin{document}
\maketitle

\begin{abstract}
Teukolsky's decoupled equations and Wald's adjoint method of metric reconstruction are built from second-order operators: the linearised Einstein operator $\mathcal G$, the curvature-extraction operator $\mathcal T$, the decoupled operator $\mathcal O$ and the reconstruction operator $\mathcal S^*$, related by $\mathcal S\mathcal G=\mathcal O\mathcal T$. We show that on four-dimensional constant-curvature backgrounds with $\Lambda\neq0$ all of these operators factor into first-order maps through a single intermediate variable: the perturbation of the self-dual half of the spin connection. We also show that for $\Lambda\neq0$ the metric perturbation is itself obtained from the connection perturbation by a first-order map. The connection therefore gives a complete description of linearised gravity, and the fourth-order identity $\mathcal S\mathcal G=\mathcal O\mathcal T$ follows by composition from a single second-order identity. Using conformal flatness, we choose the Hertz-potential map so that, after a conformal rescaling, the potential obeys the flat wave equation component by component. This leads to a new reconstruction scheme, in which the potential is sourced by the curvature $\psi$, which in turn obeys a decoupled sourced equation. A prescribed curvature is then reached by a second-order rather than a fourth-order equation, and the corrector is local and algebraic in the stress-energy tensor. The price is that the full curvature, rather than a single component, must be determined, and that the reality of the reconstructed metric becomes a condition, which we argue holds for retarded solutions. We illustrate the scheme for a static point mass in de Sitter space, where it gives a real metric perturbation that agrees with linearised Schwarzschild--de Sitter up to a diffeomorphism. We view these results as a proof of concept for a first-order Teukolsky--Wald framework on Kerr.
\end{abstract}

\section{Introduction}

Perturbation theory provides a principal means of studying gravitational fields beyond the relatively small class of explicitly known solutions of Einstein's equations. Even at the linearised level, these equations form a coupled system whose geometric structure and physical content can be difficult to extract. The study of perturbations of the Schwarzschild and Kerr solutions provides particularly important examples~\cite{TheMathematicaChandr1998,Perturbations_o_Teukol_1973}. A central idea is to replace the coupled tensor equations by simpler equations for scalar quantities, and then reconstruct the original fields by applying differential operators. Such constructions occur in the Hertz and Debye potential descriptions of Maxwell fields~\cite{ElectromagneticCohen1974}, their Yang--Mills generalisations~\cite{Debye_Potential_Castil_1999}, and treatments of higher-spin and half-integer-spin fields~\cite{ElectromagneticDolan2019,HertzAndDebyeMustaf1987,Symmetry_operat_Araned_2016}. For gravity, scalar potentials are important both in black-hole perturbation theory and metric reconstruction~\cite{Black_hole_pert_Pound_2021}, and in the study of amplitudes and correlators in self-dual settings~\cite{Pure_connection_Krasno_2024,A_double_copy_f_Campig_2021}.

Two properties make a scalar description especially useful: decoupling, whereby scalar components satisfy independent equations, and separability, whereby a suitable coordinate ansatz reduces these equations to ordinary differential equations. The Teukolsky equation on Kerr is the paradigmatic example~\cite{Perturbations_o_Teukol_1973}. Deriving such an equation for a curvature component and reconstructing a metric perturbation from a scalar potential are related but distinct problems. Wald's adjoint-operator interpretation places them in a common framework~\cite{Wald:1978vm}. The derivation of a decoupled equation is expressed as an operator identity
\[
    \mathcal{S}\mathcal{G}=\mathcal{O}\mathcal{T},
\]
where $\mathcal{G}$ is the linearised Einstein operator, $\mathcal{T}$ extracts a curvature quantity from a metric perturbation, and $\mathcal{O}$ is its decoupled operator. Using the formal self-adjointness of $\mathcal{G}$, the adjoint identity is
\begin{equation}\label{teuk-adj-intr}
    \mathcal{G}\mathcal{S}^*=\mathcal{T}^*\mathcal{O}^*.
\end{equation}
It gives the reconstruction prescription
\[
    \mathcal{O}^*\phi=0
    \quad\Longrightarrow\quad
    h=\mathcal{S}^*\phi,\qquad \mathcal{G}h=0.
\]
Thus $\mathcal{T}$ extracts curvature from a metric perturbation, whereas $\mathcal{S}^*$ reconstructs a metric perturbation from a Hertz potential satisfying the adjoint equation. The same operator identity also governs decoupling and reconstruction with sources, but this has subtleties, and will be discussed below. 

The question addressed in this paper is whether the familiar second-order decoupling and reconstruction framework, including its sourced extension, can be understood in terms of first-order maps. This turns out to be possible using the chiral formalism for General Relativity (GR), and we develop such a description on four-dimensional constant-curvature backgrounds, with a view to an eventual treatment of Kerr. Our result is a proof of concept for a first-order version of the Teukolsky--Wald framework; we hope to extend this to Kerr in subsequent publications. The key idea is to use the chiral connection perturbation as the intermediate variable: curvature extraction and metric reconstruction then each proceed through first-order steps. Our results are of three kinds. First, on constant-curvature backgrounds with $\Lambda\neq0$, all second-order operators of the Teukolsky--Wald construction factor into first-order maps through the connection, and the Teukolsky--Wald identity follows from a single second-order identity. Second, conformal flatness allows a choice of reconstruction map for which every component of the rescaled potential obeys the flat wave equation. Third, the first-order structure leads to a new sourced reconstruction scheme, in which the potential is sourced by the curvature and the corrector is local; we test it on a static point mass in de Sitter space.

Before we describe our construction, it is useful to recall the situation in electromagnetism. For a Maxwell potential $A$ with field strength $F=dA$, Maxwell's equations read $d^* d A=0$, where $d^*$ is the adjoint of the exterior derivative operator $d$. Thus, the Maxwell operator $d^* d$ is factorised into the product of $d$ and its adjoint. The purpose of this paper is to describe a similar factorisation of the Einstein operator. 

To exhibit it, we first need the chiral version of the Maxwell factorisation, as it is this version that admits the gravitational analogue. Let $F^+$ denote the self-dual part of $F=dA$. Since $d^*\star F$ is proportional to $\star\,ddA=0$, the Maxwell operator involves only half of the field strength,
\[
d^*dA=2\,d^*F^+,
\]
so that Maxwell's equations are equivalent to $d^*F^+=0$. In terms of the first-order map, defining $d_+:=P^+d$, where $P^+$ projects onto self-dual two-forms, the Maxwell operator is, up to normalisation, the composition $d_+^*d_+$. This description of the Maxwell operator also gives a reconstruction mechanism via Hertz potentials. For a self-dual two-form $\phi$, the potential $A=d_+^*\phi$ solves Maxwell's equations whenever $d_+d_+^*\phi=0$, which on flat space is the wave equation acting on each component of $\phi$.

The first of our results is that linearised gravity with $\Lambda\neq0$ has exactly this structure. This structure only becomes visible in the chiral formalism for GR, which uses geometric objects adapted to self-dual/anti-self-dual decomposition, similar to what was described in the preceding paragraph for Maxwell theory. In the chiral perturbation theory for GR, the perturbation $a$ of the self-dual (chiral) half of the spin connection plays the role of $A$, and the linearised self-dual Weyl curvature $\psi=d_2a$ plays the role of $F^+$. The operator $d_2$ is defined below. The identity \eqref{identity-intr} below states that the linearised Einstein operator on connection perturbations is $E_a=d_2^*d_2$, the analogue of the Maxwell operator $d_+^*d_+$. This also gives a Hertz potential reconstruction mechanism. For a potential $\phi$, the connection $a=d_2^* \phi$ satisfies the connection Einstein equation $E_a a=0$ whenever $d_2 d_2^* \phi=0$.  

The new ingredient compared with electromagnetism is the metric. For $\Lambda\neq0$, the Einstein condition for the curvature of the self-dual connection can be read backwards, and the metric can be extracted from the curvature of the self-dual connection. At the linearised level this gives a first-order map $\sigma=d_1^*a$ from the connection perturbation back to the metric perturbation, where $d_1$ is a first-order differential operator defined below, and $d_1^*$ is its adjoint. This has two implications. First, there is an equivalent description of Einstein perturbations via connections. Second, every second-order operator of the Teukolsky--Wald construction factors into first-order maps through the connection; see below.

We now describe this construction, including its implications for sourced reconstruction, in more detail. Given that the chiral formalism for GR is not universally known, we start with a quick roadmap. The main line of argument uses only three facts about the chiral formulation of GR; see Section~\ref{sec:pelb-form-of-gr} for details.
\begin{enumerate}
\item The metric is encoded in a triple of self-dual two-forms $\Sigma^i$, $i=1,2,3$, subject to $\Sigma^i\wedge\Sigma^j\sim\delta^{ij}$, from which it is recovered algebraically. In terms of a null tetrad, the $\Sigma^i$ are, up to normalisation, the three self-dual bivectors. The perturbation $\sigma^i=\delta\Sigma^i$ is an algebraic repackaging of the metric perturbation together with three internal rotations: $\sigma^i_{\mu\nu}=\delta g_{[\mu}{}^\rho\Sigma^i_{|\rho|\nu]}+\epsilon^{ijk}\omega^j\Sigma^k_{\mu\nu}$, and $\delta g$ is recovered from $\sigma$ by the algebraic map $P_{10}$ of Section~\ref{sec:conformal-reconstruction}.
\item The connection $A^i$ is the self-dual half of the Levi-Civita spin connection. It is determined algebraically by the first derivatives of $\Sigma$ through $d^A\Sigma^i=0$, just as the torsion-free condition determines the spin connection from a tetrad.
\item The curvature $F^i=dA^i+\frac12\epsilon^{ijk}A^j\wedge A^k$ contains the self-dual Weyl curvature, the scalar curvature and the tracefree Ricci tensor. The Einstein equations with cosmological constant are $F^i=\Psi^{ij}\Sigma^j+\frac\Lambda3\Sigma^i$, where the symmetric tracefree matrix $\Psi^{ij}$ is the self-dual Weyl curvature.
\end{enumerate}
For the benefit of the reader, Table~\ref{tab:dictionary} translates the main objects into metric, two-spinor and Newman--Penrose language.

\begin{table}[ht]
\centering
\renewcommand{\arraystretch}{1.3}
\begin{tabular}{p{0.3\textwidth}p{0.62\textwidth}}
\hline
This paper & Metric, two-spinor and Newman--Penrose language\\
\hline
$\Sigma^i$ & metric; the self-dual bivectors $\sigma^i_{AB}\,\epsilon_{A'B'}\,\theta^{AA'}\wedge\theta^{BB'}$ of a null tetrad\\
$\sigma^i=\delta\Sigma^i$ & $\delta g_{\mu\nu}$ together with three internal rotations; $\delta g=P_{10}(\sigma)$\\
$A^i$, $a^i$ & self-dual (unprimed) spin connection and its perturbation; the twelve complex NP spin coefficients\\
$\Psi^{ij}$, $\psi^{ij}$ & Weyl spinor $\Psi_{ABCD}$ and its perturbation $\psi_{ABCD}$; the Weyl scalars $\psi_0,\dots,\psi_4$\\
$\psi=d_2a$ & linearised Weyl tensor components from the connection\\
$d_2^*\psi=0$ & Bianchi identity $\nabla^{AA'}\psi_{ABCD}=0$\\
$E_\sigma\sigma=0$, $E_aa=0$ & linearised Einstein equations $\delta(R_{\mu\nu}-\Lambda g_{\mu\nu})=0$\\
$\phi$, $d_3$ & Hertz potential, and the first factor of the reconstruction operator $\mathcal S^*$\\
rescaled $\mathcal P\propto\hat\Box$ & conformal invariance of the massless spin-two equation~\cite{PenroseRindler1984}\\
\hline
\end{tabular}
\caption{Dictionary between the objects used in this paper and their familiar counterparts.}
\label{tab:dictionary}
\end{table}

We now explain how the operators $d_1, d_2$ and their adjoints arise. Linearising the equation $d^A \Sigma^i=0$ defines a first-order map from the 2-form perturbation to the connection perturbation,
\[
    a=\delta A=d_1\sigma.
\]
The linearised Einstein condition supplies the curvature equation
\[
    d^A a^i=\psi^{ij}\Sigma^j+\frac{\Lambda}{3}\sigma^i,
\]
where $\psi^{ij}$ is the symmetric tracefree matrix representing the linearised chiral Weyl curvature. Its two algebraically distinct parts define two further first-order maps. Let $P_5$ project onto the five-dimensional space of two-forms of the form $\psi^{ij}\Sigma^j$. Linearisation of the $\Sigma^i\wedge \Sigma^j\sim \delta^{ij}$ condition gives $P_5\sigma=0$, see below. This means that the Weyl projection defines curvature extraction, while the complementary projection gives, for $\Lambda\neq0$, the reverse map to the 2-form perturbation:
\[
    d_2a:=P_5d^A a=\psi^{ij}\Sigma^j,
    \qquad
    d_1^*a:=\frac{3}{\Lambda}(\mathbb{I}-P_5)d^A a=\sigma.
\]
The latter is the formal adjoint of $d_1$ for the pairings specified in the main text. We emphasise that it is the invertibility of the matrix $\Psi+\frac{\Lambda}{3}\mathbb{I}$ appearing in the curvature equation that makes the map $\sigma= d_1^* a$ possible. On constant curvature backgrounds considered here this is equivalent to $\Lambda\neq0$, and this non-zero constant is absorbed into the normalisation of $d_1^*$. Identifying $\psi^{ij}\Sigma^j$ with its coefficient matrix, the forward maps are summarised by
\begin{equation}\label{diagram-intr}
\begin{tikzcd}[column sep=large]
    \sigma \arrow[r, "d_1"] & a \arrow[r, "d_2"] & \psi .
\end{tikzcd}
\end{equation}

The linearised Einstein equations are therefore the first-order system 
\[
a=d_1\sigma, \qquad \sigma=d_1^*a. 
\]
Eliminating the connection gives
\[
    E_\sigma\sigma=0,\qquad E_\sigma:=d_1^*d_1-\mathbb{I}.
\]
Alternatively, eliminating the 2-form perturbation gives the pure connection equation
\[
    E_a a=0,\qquad E_a:=d_1d_1^*-\mathbb{I}.
\]
The two descriptions are equivalent. If $E_a a=0$, then $\sigma=d_1^*a$ satisfies $E_\sigma\sigma=0$ by the identity $E_\sigma d_1^*=d_1^*E_a$, and its compatible connection is $d_1\sigma=a$. Conversely, if $E_\sigma\sigma=0$, then $a=d_1\sigma$ satisfies $E_a a=0$, and $d_1^*a=\sigma$. Thus $d_1$ and $d_1^*$ are mutually inverse on the solution spaces: every linearised Einstein solution can be described by a connection perturbation and reconstructed from it. This pure connection description of the gravitational perturbation theory on a constant curvature background is not new, see \cite{Delfino:2012zy}, \cite{Delfino:2012aj}, but the interpretation in terms of first-order operators developed here is new. We note that the fact that there is a pure connection version of linearised GR is not surprising in view of the fact that there is a fully non-linear pure connection formalism \cite{Krasnov:2011pp}. The equation $E_a a=0$ is the linearisation of the field equations in this pure connection formalism. 

The next question is whether the operator diagram (\ref{diagram-intr}) extends this equivalence to a description in terms of curvature alone. Our principal operator identity, proved in Section~\ref{subsec:curvature-identity}, is
\begin{equation}\label{identity-intr}
    d_1d_1^*-\mathbb{I}=d_2^*d_2,
\end{equation}
where $d_2^*$ is the formal adjoint for the pairings specified below. This identity equates two compositions of first-order maps on connection perturbations, with the zeroth-order term arising from the background curvature. This is the key identity from which the first-order version of the Teukolsky--Wald identity follows by further composition, as we now explain. 

For an Einstein solution, its curvature $\psi=d_2a$ consequently obeys
\[
    d_2^*\psi=0,
    \qquad\text{and hence}\qquad
    E_\psi\psi=0,\qquad E_\psi:=d_2d_2^*.
\]
This equation is of course the differential Bianchi identity, written in the language of operator $d_2$ and its adjoint. Here the analogy with the 2-form--connection correspondence stops. There is no subtraction of $\mathbb{I}$ in $E_\psi$, and $a$ cannot be recovered from $\psi$: applying $d_2^*$ to the actual curvature gives zero. Thus the identity (\ref{identity-intr}) does not provide an equivalent "pure curvature" description of the full solution space. 

The same identity nevertheless supplies a solution-generating procedure. Let $\phi$ be a curvature-type potential satisfying $E_\psi\phi=0$. Then $a=d_2^*\phi$ obeys
\[
    E_a a=d_2^*E_\psi\phi=0,
\]
and $\sigma=d_1^*a$ gives a metric perturbation. This is a construction from a potential, rather than an inverse to curvature extraction: its output has $d_2a=0$, and potentials in $\ker d_2^*$ produce the zero connection. 

The next step uses the fact that the background geometry is conformally flat. To motivate it, we note that one can replace $d_2^*$ in $E_\psi=d_2 d_2^*$ above by a conveniently chosen first-order map $d_3$, also mapping "curvatures" to "connections", and set
\[
    \mathcal{P}:=d_2d_3,\qquad E_a d_3=d_2^*\mathcal{P}.
\]
Every solution of $\mathcal{P}\phi=0$ then generates an Einstein solution through
\[
    a=d_3\phi,\qquad \sigma=d_1^*a,
\]
followed by the algebraic extraction of the metric from $\sigma$. The construction uses two first-order differential steps and generates the sector with vanishing linearised chiral Weyl curvature, since $d_2a=\mathcal{P}\phi=0$. As we discuss in detail below, see Section~\ref{subsec:version-I}, this is the analogue of Wald's standard construction. Its output has a single helicity and is therefore complex, like the metric perturbation produced by Wald's construction before its real part is taken.

The freedom to choose $d_3$ allows us to simplify the equation for the potential. The natural choice $d_3=d_2^*$ gives a gauge-covariant wave operator whose connection terms couple its components. On constant-curvature backgrounds for which $\hat g=\lambda^2g$ is the flat metric, we use conformal flatness to choose a weighted adjoint that removes these couplings. Thus, it turns out that the best choice is $d_3=wd_2^*$, where $w=\lambda^{-4}$. With this choice, and after rescaling the potential as $\phi=\lambda^3\chi$, its equation becomes the flat wave equation acting independently on each component
\[
\mathcal{P} (\lambda^3 \chi) = - \frac{3}{2\Lambda} \lambda \hat{\Box} \chi,
\]
where $\hat{\Box}$ is the flat box operator. Thus the operator identity establishes the reconstruction mechanism, while conformal geometry selects a particularly useful realisation of it. The fact that, after rescaling, the same flat wave operator also governs the linearised chiral Weyl curvature reflects the conformal invariance of the massless spin-2 equation on conformally flat backgrounds~\cite{PenroseRindler1984}. What is new here is the first-order factorisation of this structure and the reconstruction map built on it.

The use of conformal geometry here has precedents. The role of conformally K\"ahler structures in simplifying field equations on algebraically special backgrounds has been explored in~\cite{SelfDualKahleDerdzi1983,Kerr_metric_fro_Krasno_2024,Symmetry_operat_Araned_2016}. Our previous work on Maxwell fields~\cite{Kahler_decoupli_Green_2026} combined this conformal viewpoint with the chiral description to elucidate decoupling. For gravity, our study of the \pleb{} complex~\cite{Plebanski_compl_Krasno_2025} expressed the linearised Einstein operator on hyper-K\"ahler backgrounds in terms of a first-order map and its adjoint. Here we similarly use conformal flatness to simplify the resulting solution-generating problem.

The complete construction is summarised by the following diagram. The horizontal path describes curvature extraction from a 2-form perturbation, while the path from $\phi$ through $a$ to $\sigma$ describes reconstruction from a potential. Although $\phi$ and $\psi$ have the same algebraic tensor type, we display them as separate nodes to distinguish these roles. The curved arrows denote formal adjoints. All arrows are first-order differential operators; the only remaining step, the extraction of $\delta g$ from $\sigma$, is algebraic.
\[
\begin{tikzcd}[column sep=huge, row sep=large]
    & \phi \arrow[d, "d_3"] & \\
    \sigma
    \arrow[r, "d_1"]
    &
    a
    \arrow[l, bend left=35, "d_1^*"]
    \arrow[r, "d_2"]
    &
    \psi
    \arrow[l, bend left=35, "d_2^*"]
\end{tikzcd}
\]

The correspondence with the adjoint Teukolsky--Wald identity can now be made precise. In the 2-form description, the Einstein operator is $\mathcal{G}=E_\sigma$, and the second-order extraction and reconstruction operators factor as
\[
    \mathcal{T}=d_2d_1,\qquad
    \mathcal{S}^*=d_1^*d_3,
    \qquad\mathcal{T}^*=d_1^*d_2^*.
\]
Our potential operator is therefore $\mathcal{P}=d_2d_3=\mathcal{O}^*$. With these identifications, the reconstruction identity reads
\[
    \underbrace{E_\sigma}_{\mathcal{G}}\,
    \underbrace{d_1^*d_3}_{\mathcal{S}^*}
    =
    \underbrace{d_1^*d_2^*}_{\mathcal{T}^*}\,
    \underbrace{d_2d_3}_{\mathcal{P}=\mathcal{O}^*},
\]
which is precisely the adjoint form displayed in (\ref{teuk-adj-intr}). Note that the Teukolsky--Wald identity is of fourth order, whereas \eqref{identity-intr} is a second-order identity from which it follows by composition.

Taking its adjoint gives the corresponding forward identity, with $\mathcal{S}=d_3^*d_1$ and $\mathcal{O}=\mathcal{P}^*=d_3^*d_2^*$. Consequently, $\psi=\mathcal{T}\sigma$ satisfies the curvature equation $\mathcal{O}\psi=0$, while a solution of the potential equation $\mathcal{P}\phi=0$ generates $\sigma=\mathcal{S}^*\phi$. The metric perturbation is then extracted algebraically from $\sigma$. This identifies $\phi$ with the Hertz-potential variable in Wald's construction. 

Let us now describe the extension of our first-order formalism to nonzero sources. As discussed in detail in Section~\ref{sec:discussion}, the first-order formalism admits two versions of sourced reconstruction. Version~I is the exact analogue of the standard Wald procedure. Version~II is a new scheme. Here we describe this new scheme, postponing the discussion of both schemes and their comparison with the standard scheme to Section~\ref{sec:discussion}.

For a conserved stress-energy tensor, let $j$ denote its algebraic image in the space of 2-form perturbations, with the normalisation specified in Section~\ref{sec:sourced-equations}. The 2-form and connection equations become
\[
    E_\sigma\sigma=j,\qquad E_a a=d_1j,
    \qquad a=d_1\sigma,\qquad \sigma=d_1^*a-j.
\]
Thus the reverse map from connections to 2-forms acquires an algebraic source correction, while curvature extraction remains $\psi=d_2a$. The 2-form and connection descriptions are still equivalent even with sources. Indeed, if a connection satisfies its field equation $E_a a = d_1 j$, then $d_1\sigma =a$, and the 2-form perturbation $\sigma = d_1^* a - j$ satisfies the Einstein equations $E_\sigma \sigma = j$, as is easy to check. 

The principal identity \eqref{identity-intr} is an identity between background operators, so it holds irrespective of sources. Applied to $\psi=d_2a$ with $E_aa=d_1j$, it gives the sourced first-order Bianchi equation $d_2^*\psi=d_1j$. For the conformally adapted choice $d_3=wd_2^*$, where $w=\lambda^{-4}$, the potential operator is then $\mathcal{P}= d_2 d_3= d_2 w d_2^*$, and $\mathcal O=\mathcal{P}^*=\mathcal P$ because $d_2wd_2^*$ is formally self-adjoint.  The second-order consequence of the sourced first-order Bianchi equation is then
\begin{equation}\label{sourced-intr}
    \mathcal O \psi=\mathcal S j.
\end{equation}
This is the sourced Teukolsky--Wald equation, with the decoupling map factorised as $\mathcal S=d_2wd_1$. The conformal rescaling reduces it to componentwise flat wave equations with differentiated stress-energy on the right-hand side, subject to the first-order Bianchi constraints.

The story so far is analogous to the second-order Teukolsky--Wald story with sources. The differences arise in the version~II reconstruction, which now proceeds in two stages. We first determine the curvature $\psi$ from the sourced curvature system $ \mathcal P\psi=\mathcal S j$, subject to the first-order constraint $d_2^*\psi = d_1 j$. We then search for a potential $\phi$ for the connection so that $a=d_3 \phi$, with the curvature of this connection to be given by $\psi$. We thus impose $\mathcal{P} \phi =\psi$. After solving for the potential $\phi$ with the curvature $\psi$ as its source, the connection and metric are recovered by first-order operators:
\[
    \mathcal P\phi=\psi
    \quad\longrightarrow\quad
    a=d_3\phi
    \quad\longrightarrow\quad
    \sigma=d_1^*a-j.
\]
Indeed, $E_a a=d_2^*\psi=d_1j$, so the reconstructed connection satisfies the sourced Einstein equations. The 2-form perturbation is then recovered from the connection perturbation. Unlike the homogeneous potential ansatz, this construction reproduces the prescribed, generally nonzero chiral curvature. This is version~II of Section~\ref{subsec:version-II}: the potential equation is given a source, as in the sourced form of Wald's construction, but the key new point is that the source is chosen to be the curvature $\psi$ itself. 

Let us compare the version~II sourced reconstruction procedure arising in our framework to the one in the standard description. Writing $J$ for a conserved stress-energy source, with its coupling constant absorbed, one has
\[
    \mathcal{G}h=J
    \quad\Longrightarrow\quad
    \mathcal{O}\psi=\mathcal{S}J,\qquad \psi=\mathcal{T}h.
\]
This is a sourced curvature equation, and our \eqref{sourced-intr} is its precise analogue. In the usual framework one then uses the adjoint identity $\mathcal{G} \mathcal{S}^* = \mathcal{T}^* \mathcal{O}^*$. Setting $h= \mathcal{S}^* \phi$ makes the metric perturbation satisfy $\mathcal{G} h = \mathcal{T}^* \mathcal{O}^* \phi$. The problem with this reconstruction is that not every source $J$ is of the form $\mathcal T^*\eta$, with $\eta=\mathcal O^*\phi$, so that a corrector must be added~\cite{Green:2020sourced}; for a point particle it is nonlocal. 

The same obstruction arises in our framework if the potential equation is given a general source. It is avoided by the choice made above, in which the source of the potential equation is the curvature $\psi$ itself. The corrector is then the local term $-j$ in $\sigma=d_1^*a-j$, and the metric automatically satisfies the linearised Einstein equation with the specified source. The same choice means that a prescribed curvature is reached by the second-order equation $\mathcal P\phi=\psi$, a componentwise flat wave equation, rather than by the fourth-order inversion relating the curvature to the Hertz potential in the standard construction. The price is twofold. First, a generic curvature, such as the Coulomb field of a point mass, is not aligned with a fixed null direction. Version~II therefore needs the full curvature $\psi$, which must satisfy the first-order sourced Bianchi equation, and a potential with all five components; on the backgrounds considered here each component obeys a decoupled flat wave equation. Second, the reconstructed perturbation is complex, and reality is a condition on the solution of $\mathcal P\phi=\psi$; we argue that it is satisfied by retarded solutions. The new sourced reconstruction mechanism we propose is illustrated by the example of a point mass in de Sitter space in Section~\ref{sec:example-static-mass}. There the reconstructed metric is real without any correction, and it agrees with the linearised Schwarzschild--de Sitter metric up to a diffeomorphism that is regular away from the worldline. Section~\ref{sec:discussion} compares the two constructions in detail and summarises what is simplified.

Most of our results are about the $\Lambda\not=0$ case, but some analogous statements can be made about $\Lambda=0$ gravity. There is no longer an equivalent connection description on the flat background, but suitable first-order maps nevertheless yield an analogous solution-generating construction. Since the message there is rather different, this case is treated in Appendix~\ref{sec:lin-eom-lambda-0}.

The paper is organised as follows. Section~\ref{sec:backgr} recalls \pleb{}'s chiral formulation and summarises the few algebraic facts used later. Section~\ref{sec:linear-eom} derives the linearised equations for $\Lambda\neq0$, establishes the equivalence of the 2-form and connection descriptions, and proves the principal operator identity. Section~\ref{sec:conformal-reconstruction} uses conformal flatness to choose the potential operator, and develops the Teukolsky--Wald comparison and the metric reconstruction. Here we also work out the construction explicitly on de Sitter space, including the resulting metric perturbation. Section~\ref{sec:sourced-equations} extends the first-order equations and the reconstruction to conserved stress-energy sources. Section~\ref{sec:example-static-mass} carries out version~II for a static point mass in de Sitter space and compares the result with linearised Schwarzschild--de Sitter. Section~\ref{sec:discussion} compares the first-order reconstruction with the standard Teukolsky--Wald construction and its extension to sources by Green, Hollands and Zimmerman (GHZ)~\cite{Green:2020sourced}, from three angles: prescribed curvature, sources and reality. It identifies precisely what is simplified, and discusses gauge, completeness and the potential extension to Kerr. The appendices collect a discussion of the reality conditions, the algebraic decompositions, the computation of the adjoints, the derivation of the connection, the action-based pairings, the case $\Lambda=0$, and the Weyl curvature of the Hertz metric.

\section{Background Material}\label{sec:backgr}

The purpose of this section is to give a brief introduction to the chiral (\pleb) formulation of GR, and to summarise the algebraic facts about $\C^3$-valued differential forms used in the later exposition. For more details about the formalism we refer the reader to \cite{Krasnov:2010olp} and \cite{Bhoja:2024xbe}, and also \cite{FormGenRelGravity2020}.

\subsection{\pleb{}'s Formulation of GR}\label{sec:pelb-form-of-gr}

This chiral formulation replaces the metric by a nondegenerate triple of two-forms $\sd^i$ satisfying an algebraic metricity constraint, see below. These forms span the self-dual subspace $\Lambda^+$ of the metric they determine. The remaining variables are an $\so(3,\C)$ connection $A^i$ and a symmetric tracefree matrix $\Psi^{ij}$ representing the chiral Weyl curvature. We identify the adjoint representation with $E=\C^3$, using $\delta^{ij}$ and $\epsilon^{ijk}$ as its invariant tensors.\footnote{Latin indices $i,j,k$ refer to the three-dimensional internal space $E$; Greek indices refer to spacetime.} In Lorentzian signature the self-dual description is complex, even when the underlying metric is real. The corresponding real Euclidean construction uses $E=\R^3$. Internal $SO(3,\C)$ rotations change the choice of self-dual frame without changing the metric.

A useful way to see that the two-forms contain the metric is the Urbantke formula~\cite{OnIntegrabilitUrbant1984}, which reconstructs its densitised components directly from the triple:
\begin{align}
	\sqrt{-g}g_{\mu\nu} = \frac{\im}{12}\epsilon^{ijk} \tilde{\epsilon}^{\rho\sigma\alpha\beta} \sd^i_{\mu\rho} \sd^j_{\nu\sigma} \sd^k_{\alpha\beta}.
\end{align}

We use Lorentzian signature $(-,+,+,+)$ and choose the orientation so that $\epsilon_{0123}=\sqrt{-g}$ in an oriented chart. The upper alternating density in the Urbantke formula is $\tilde\epsilon^{\mu\nu\rho\sigma}=\sqrt{-g}\,\epsilon^{\mu\nu\rho\sigma}$, so $\tilde\epsilon^{0123}=-1$. Our Hodge convention is
\[
 (\star\alpha)_{\mu_{p+1}\ldots\mu_4}
 =\frac1{p!}\alpha^{\mu_1\ldots\mu_p}\epsilon_{\mu_1\ldots\mu_4},
 \qquad \star\sd^i=\im\sd^i,
 \qquad \star^2\big|_{\Lambda^p}=(-1)^{p(4-p)+1}.
\]
The bilinear metric pairing of $p$-forms includes the factor $1/p!$. In particular, $(B,C)=\frac12 B_{\mu\nu}C^{\mu\nu}$ for two-forms. These conventions imply $\sd^{i\mu\nu}\sd^j_{\mu\nu}=4\delta^{ij}$.

The metricity, compatibility, and curvature equations are, respectively,
\begin{gather}
	\sd^i \wedge \sd^j \sim \delta^{ij}, \label{eq:metricity-cond} \\
	d^A \sd^i = d \sd^i + \epsilon^{ijk} A^j \wedge \sd^k = 0, \label{eq:torsion-free}\\
	F^i = \Psi^{ij} \sd^j + \frac{\Lambda}{3} \sd^i, \label{eq:ee-curv-cond}
\end{gather}

Here $\sim$ means proportionality by a common four-form, $d^A$ is the gauge-covariant exterior derivative, and $F^i=dA^i+\frac12\epsilon^{ijk}A^j\wedge A^k$ is the curvature. The metricity condition says that the wedge-product matrix is pure trace. It removes five of the eighteen components of an arbitrary triple of two-forms, leaving the ten metric components and three internal frame rotations. 

The compatibility equation determines $A$ from the first derivatives of $\sd$. This is the chiral analogue of determining the torsion-free spin connection from a tetrad. Its solution uses the algebraic endomorphism $J_1$ of $E$-valued one-forms, defined by $J_1(\theta)^i_\mu=\epsilon^{ijk}\sd^j_\mu{}^\nu\theta^k_\nu$. Applying the Hodge star $\star:\Lambda^k\to\Lambda^{4-k}$ gives
\begin{align}
	A^i = \im J_1^{-1}(\star d\sd^i). \label{eq:con-def}
\end{align}

Appendix~\ref{apx:torsion-free-sol} derives this formula. Once $A$ is the compatible connection, the curvature equation imposes the Einstein condition: the mixed self-dual/anti-self-dual curvature block vanishes, while the trace of the self-dual block is fixed by $\Lambda$. Its symmetric tracefree part is $\Psi$, which remains dynamical. With the convention in \cref{eq:ee-curv-cond}, $\operatorname{Ric}=\Lambda g$ and the scalar curvature is $4\Lambda$; see~\cite{FormGenRelGravity2020}.

The algebraic equations can first be studied over the complex numbers. To obtain a real Lorentzian metric one imposes reality conditions, including the orthogonality of the triple to its complex conjugate,
\begin{align}
	\sd^i \wedge \asd^j = 0.
\end{align}
We derive the consequences of these reality conditions for 2-form perturbations in Appendix~\ref{apx:metric-reality}. Here $\asd^i$ denotes the conjugate anti-self-dual triple for a real Lorentzian metric. The volume normalisation must also be chosen consistently with reality. With the normalisation used in this paper, raising one spacetime index turns each self-dual two-form into an endomorphism, and these endomorphisms satisfy the quaternion algebra
\begin{align}\label{eq:quat-alg}
	\sd^i_\mu{}^\rho\ \sd^j_\rho{}^\nu = -\delta^{ij} \delta_\mu{}^\nu + \epsilon^{ijk} \sd^k_\mu{}^\nu.
\end{align}

We will specialise to constant-curvature backgrounds, for which both chiral parts of the Weyl tensor vanish. In particular $\Psi=0$, and the curvature of the background connection obeys the relation below. This relation constrains only the self-dual curvature block: it characterises Einstein backgrounds with vanishing self-dual Weyl curvature. The stronger assumption of constant curvature, with both halves of the Weyl tensor vanishing, is used for the conformal reduction later in the paper.
\begin{align}
	F^i = \frac{\Lambda}{3} \sd^i.
\end{align}

\subsection{Algebraic Facts Used Below}\label{subsec:algebra-summary}

The first-order maps introduced in the next section act on $E$-valued one-forms and two-forms. We use only the following pointwise algebraic facts, which depend on the background triple $\sd$ and involve no equations of motion; their derivation is given in Appendix~\ref{sec:decomp-of-diff-forms}.
\begin{itemize}
\item $E$-valued one-forms, which have twelve components, split as $E\otimes\Lambda^1=(E\otimes\Lambda^1)_4\oplus(E\otimes\Lambda^1)_8$, the eigenspaces of the algebraic operator $J_1$ with eigenvalues $2$ and $-1$. Hence $J_1^{-1}=\frac12(J_1-\mathbb I)$. The four-dimensional summand consists of the elements $\sd^i\cdot\xi$, with $(\sd^i\cdot\xi)_\mu=\sd^i_\mu{}^\nu\xi_\nu$, built from ordinary one-forms $\xi$.
\item $E$-valued two-forms, which have eighteen components, split as $1\oplus3\oplus5\oplus9$, with representatives $h\sd^i$, $\epsilon^{ijk}h^k\sd^j$, $h^{ij}\sd^j$ with $h^{ij}$ symmetric and tracefree, and $\theta^{ij}\asd^j$, where $\asd^i$ is the anti-self-dual triple. The $1\oplus9$ part encodes a metric perturbation, the three-dimensional part internal rotations, and the five-dimensional part has the algebraic type of the self-dual Weyl curvature.
\item $P_5$ denotes the projection onto the five-dimensional summand. A linearised 2-form perturbation obeys $P_5\sigma=0$, whereas the linearised chiral Weyl curvature lies in the image of $P_5$. When convenient, we identify $\psi^{ij}\sd^j$ with its coefficient matrix $\psi^{ij}$.
\end{itemize}

\section{Linearised Equations of Motion}\label{sec:linear-eom}

We now linearise about a constant-curvature background with $\Lambda\neq0$. This section establishes the first-order maps, the equivalence of the 2-form and connection equations, and the curvature-dependent operator identity. Background fields retain the notation $\sd,A,\Psi$, and their perturbations are introduced by
\begin{align}
	\sd^i \rightarrow \sd^i + \sigma^i, \quad A^i \rightarrow A^i + a^i, \quad \Psi^{ij} \rightarrow \Psi^{ij} + \psi^{ij}
\end{align}
where $\sigma^i,a^i,\psi^{ij}$ are the perturbations and the background has $\Psi=0$. The curvature expansion is
\begin{align}
	F^i \rightarrow F^i + d^A a^i + \frac12\epsilon^{ijk} a^j\wedge a^k
\end{align}
Only the term $d^Aa$ contributes to the linearised curvature; the final term is quadratic in the perturbation and is not used below.

\subsection{First-Order Maps and Their Adjoints}

Linearising metricity, compatibility, and the Einstein curvature equation in \cref{eq:metricity-cond,eq:torsion-free,eq:ee-curv-cond} gives
\begin{align}
	\sd^{(i} \wedge \sigma^{j)} \sim \delta^{ij}, \label{eq:lin-metricity-cond}\\
	d^A \sigma^i + \epsilon^{ijk} a^j \wedge \sd^k = 0, \label{eq:lin-torsion-free} \\
	d^A a^i = \psi^{ij} \sd^j + \frac{\Lambda}{3} \sigma^i. \label{eq:lin-ee-curv-cond}
\end{align}
The first equation is algebraic: it sets the five-dimensional component of $\sigma$ to zero. Thus $\sigma$ takes values in the $1+3+9$ summands of \cref{subsec:2-form-irrep-decomp}. The remaining thirteen components describe the metric perturbation and the choice of internal frame. We impose this constraint from the outset; all operators acting on 2-form perturbations below have this restricted domain.

The compatibility equation \cref{eq:lin-torsion-free} determines the connection perturbation from the first derivatives of $\sigma$, exactly as its nonlinear counterpart determines $A$ from $\sd$:
\begin{align}
	a^i = \im J_1^{-1}{(\star d^A \sigma)}^i, \quad \Rightarrow \quad a^i = d_1(\sigma){}^i.
\end{align}
This defines $d_1$, the linearised compatible-connection map. Its domain is the constrained 2-form space and its target is $E\otimes\Lambda^1$.

The curvature equation \cref{eq:lin-ee-curv-cond} determines two further maps. Its Weyl term lies in the image of $P_5$, whereas $\sigma$ lies in the complementary subspace. Applying the Weyl projection therefore gives
\begin{align}
	P_5(d^A a){}^i = P_5 (\psi \sd){}^i + \frac{\Lambda}{3} P_5(\sigma){}^i, \quad \Rightarrow \quad \psi^{ij} \sd^j = d_2(a){}^i = P_5(d^A a){}^i.
\end{align}
This is the curvature-extraction map $d_2$. Applying the complementary projection and using $\Lambda\neq0$ reconstructs the 2-form perturbation from the connection:
\begin{align}
	d^A a^i = P_5(d^A a){}^i + \frac{\Lambda}{3} \sigma^i, \quad \Rightarrow \quad \sigma^i = d_1^*(a) = \frac{3}{\Lambda} (\mathbb{I} - P_5)(d^A a){}^i.
\end{align}

\subsection{Adjoints}

The reverse map is denoted by $d_1^*$ because it is the formal adjoint of $d_1$ for the following pairings:
\begin{align}
	\langle a, a \rangle = \im ( a^i, J_1(a){}^i), \\
	\langle \sigma, \sigma \rangle = \frac{\Lambda}{3}(\sigma^i, \star \sigma^i).
\end{align}
Here $(\cdot,\cdot)$ denotes the metric pairing of differential forms. Formal adjoints refer to the corresponding integrated pairings, with boundary terms discarded, for example for compactly supported fields. In the complex chiral setting these are bilinear pairings, not positive-definite Hermitian products; the adjoint notation should not be interpreted as a positivity statement. Their action-based origin is discussed in Appendix~\ref{apx:inner-products}; see also~\cite{Plebanski_compl_Krasno_2025}. The verification is a direct integration by parts, given in Appendix~\ref{apx:adjoints}.

To compute the adjoint of curvature extraction, equip its target with the pairing
\begin{align}
			\langle \psi, \psi \rangle = -\frac{3}{\Lambda}(\psi^{ij} \sd^j, \star \psi^{ik} \sd^k).
\end{align}
The overall sign is fixed by the same rule as for the connection and 2-form pairings; see Appendix~\ref{apx:inner-products}. With the opposite sign of the pairing, the operator $d_2^*$ introduced below would be minus the formal adjoint of $d_2$. With this pairing, integration by parts (Appendix~\ref{apx:adjoints}) gives the formal adjoint
\begin{align}
		d_2^*(\psi){}^i = -\frac{3\im}{\Lambda}J_1^{-1}(\star d^A (\psi \sd)){}^i.
\end{align}

\subsection{Equivalent 2-Form and Connection Equations}

We next compose the first-order maps to obtain the equations of motion. For reference, the four operators are
\begin{align}
	d_1(\sigma){}^i & = \im J_1^{-1}(\star d^A \sigma){}^i, \quad d_1^*(a){}^i = \frac{3}{\Lambda} (\mathbb{I} - P_5)(d^A a){}^i, \\
	d_2(a){}^i & = P_5(d^A a){}^i, \quad d_2^*(\psi){}^i = -\frac{3\im}{\Lambda} J_1^{-1}(\star d^A (\psi \sd)){}^i.
\end{align}
Once metricity is imposed, the compatibility and curvature equations reduce to the coupled first-order system
\begin{align}
	a^i = d_1(\sigma){}^i, \quad \sigma^i = d_1^*(a){}^i.
\end{align}

Eliminating the connection gives a second-order equation for the 2-form perturbation, and hence for the metric it encodes:
\begin{align}
	\sigma^i = (d_1^* d_1)(\sigma){}^i, \quad \Rightarrow \quad E_\sigma = d_1^* d_1 - \mathbb{I}.
\end{align}
As in the introduction, $E_\sigma$ denotes the Einstein operator on constrained two-forms. Alternatively, eliminating the 2-form perturbation gives the pure connection operator $E_a$:
\begin{align}\label{eq:lin-pure-con-ee}
	a^i = (d_1 d_1^*)(a){}^i, \quad \Rightarrow \quad E_a = d_1 d_1^* - \mathbb{I}.
\end{align}

If $a$ solves the connection equation, define $\sigma=d_1^*a$. The intertwining identity
\begin{align}
	(E_\sigma d_1^*)(a) = (d_1^* E_a)(a) = 0,
\end{align}
shows that the reconstructed 2-form perturbation solves the Einstein equation. Moreover, $d_1\sigma=a$ on a connection solution. Conversely, a 2-form solution defines $a=d_1\sigma$, which satisfies the connection equation and returns the original 2-form perturbation under $d_1^*$. Thus these maps are mutually inverse on the two solution spaces for $\Lambda\neq0$. The equivalence includes all solutions in this linearised chiral description.

\subsection{The Curvature Identity}\label{subsec:curvature-identity}

The purpose of this subsection is to derive the key identity on which our first-order approach is based. 
We have the following identity on arbitrary connection perturbations:
\begin{align}\label{eq:pure-con-wald-identity}
	d_2^* d_2 = d_1 d_1^* - \mathbb{I}.
\end{align}
The left side of the equality becomes,
\begin{align}
	(d_2^* d_2)(a){}^i = -\frac{3\im}{\Lambda} (J_1^{-1} \star d^A P_5 d^A)(a){}^i,
\end{align}
and for the right side,
\begin{align}
	(d_1 d_1^*)(a){}^i = \frac{3\im}{\Lambda} (J_1^{-1} \star d^A (\mathbb{I} - P_5) d^A )(a){}^i = \frac{3\im}{\Lambda} (J_1^{-1} \star d^A d^A)(a){}^i - \frac{3\im}{\Lambda} (J_1^{-1} \star d^A P_5 d^A)(a){}^i.
\end{align}
From here we make use of the identity
\begin{align}
	d^A d^A a^i = \epsilon^{ijk} F^j a^k = \frac{\Lambda}{3} \epsilon^{ijk} \sd^j a^k = -\frac{\im\Lambda}{3} \star J_1(a){}^i.
\end{align}
Using this with the previous line shows,
\begin{align}
	(d_1 d_1^*)(a){}^i = (J_1^{-1} \star \star J_1)(a){}^i - \frac{3\im}{\Lambda}(J_1^{-1} \star d^A P_5 d^A)(a){}^i = a^i - \frac{3\im}{\Lambda}(J_1^{-1} \star d^A P_5 d^A)(a){}^i.
\end{align}
This proves \cref{eq:pure-con-wald-identity}. The term $\mathbb I$ is the contribution of the background curvature to $(d^A)^2$; it is essential to the relation between the two compositions. 

\subsection{Potential Reconstruction}

The derived operator identity \eqref{eq:pure-con-wald-identity} allows us to formulate the reconstruction mechanism. First, using \eqref{eq:pure-con-wald-identity}, we see that any linearised connection satisfying $E_a a= (d_1 d_1^* - \mathbb{I})a=0$ also satisfies $d_2^* d_2 a=0$. Thus, for any connection satisfying $E_a a=0$, the object $\psi=d_2a$ consequently satisfies $d_2^*\psi=0$, and therefore also $d_2d_2^*\psi=0$. Unlike $d_1$ and $d_1^*$, however, $d_2$ and $d_2^*$ are not inverse on solutions: applying $d_2^*$ to the actual curvature gives zero. The identity instead allows a construction from curvature-type potentials. 

Let $\phi$ be a symmetric tracefree potential, identified with an element of $(E\otimes\Lambda^2)_5$, and choose a first-order map $d_3$ from this space to $E\otimes\Lambda^1$. If the potential satisfies the equation
\begin{align}
	\mathcal{P}(\phi) = (d_2 d_3)(\phi) = 0,
\end{align} 
its image under $d_3$ solves the connection Einstein equation. We denote the potential operator by $\mathcal P$, as in the introduction. Indeed, the operator identity gives the solution-generating implication
\begin{align}
	a^i := d_3(\phi) \quad {\rm and} \quad \mathcal{P}(\phi) = 0, \quad \Rightarrow \quad E_a (a) = (d_2^* d_2)(a) = (d_2^* d_2 d_3)(\phi) = (d_2^* \mathcal{P})(\phi) = 0.
\end{align}
The metric follows by applying $d_1^*$ and then the algebraic metric projection. The generated connection has $d_2a=0$, so $\phi$ is a potential, not its extracted chiral curvature. It remains to choose $d_3$ so that the second-order potential equation is readily solvable. This is a separate geometric problem, addressed using conformal flatness in the next section.

\subsection{A Simpler Form of the \texorpdfstring{$d_2^*$}{d2*} Operator}

The purpose of this subsection is to obtain a simpler expression for the operator $d_2^*$. We first note that the derivative of a symmetric tracefree argument produces a connection-type field in the eight-dimensional summand:
\begin{align}\label{a-8}
	J_1(\sd \cdot d^A \psi)^i = - \sd^j \cdot d^A \psi^{ij} \quad \Rightarrow \quad \sd^j \cdot d^A \psi^{ij} \in (\Eoneforms)_8.
\end{align} 
This is checked using the quaternion algebra, and it allows $J_1^{-1}$ to be eliminated from $d_2^*$; see Appendix~\ref{apx:adjoints}. This means that there is a simpler expression for the operator $d_2^*$, namely
\begin{align}\label{d2-star}
(d_2^* \psi)^i = \frac{3}{\Lambda} \sd^j \cdot d^A \psi^{ij}.
\end{align}
This will be useful in computations in the next section.

\subsection{Gauge Transformations}\label{subsec:gauge}

In this subsection we briefly describe how gauge transformations act in the chiral formalism for GR. We only need this at the linearised level for purposes of this paper. The linearised theory has two kinds of gauge symmetry: internal $\mathrm{SU}(2)$ rotations of the triple $\sd^i$, and diffeomorphisms. Both are most simply described at the level of the connection perturbation.

An infinitesimal internal rotation with parameter $\omega^i$ acts on the connection perturbation as
\begin{align}\label{eq:gauge-su2}
	\delta_\omega a^i=(d^A\omega)^i=d\omega^i+\epsilon^{ijk}A^j\omega^k .
\end{align}
An infinitesimal diffeomorphism generated by a vector field $\xi$ acts on the full connection by the Lie derivative, and Cartan's formula gives
\begin{align}
	\mathcal L_\xi A^i=i_\xi F^i+(d^A\, i_\xi A)^i .
\end{align}
The second term is an internal rotation, with parameter $i_\xi A$, and can be removed by a transformation \eqref{eq:gauge-su2}. Up to such a rotation, a diffeomorphism therefore acts on the connection perturbation algebraically,
\begin{align}\label{eq:gauge-diffeo}
	\delta_\xi a^i=i_\xi F^i=\frac{\Lambda}{3}\,i_\xi\sd^i ,
\end{align}
where we used the background equation $F^i=\frac\Lambda3\sd^i$. This is a feature of $\Lambda\neq0$: on a flat background the right-hand side vanishes. In the notation $(\sd^i\cdot\zeta)_\mu=\sd^i_\mu{}^\nu\zeta_\nu$ used in Section~\ref{sec:conformal-reconstruction}, $i_\xi\sd^i=-\sd^i\cdot\xi$, so these are the connection perturbations $a^i=\sd^i\cdot\zeta$ with $\zeta=-\frac\Lambda3\xi$. The map $\xi\mapsto i_\xi\sd^i$ is injective, and its image is the four-dimensional summand $(\Eoneforms)_4$ in the decomposition $\Eoneforms=(\Eoneforms)_8\oplus(\Eoneforms)_4$. Diffeomorphisms thus shift the four-dimensional part of $a$, while its eight-dimensional part changes only through the accompanying internal rotation.

Two consequences are worth recording. First, diffeomorphisms can be fixed algebraically, by the condition $a_4=0$ on the four-dimensional part of the connection perturbation. This condition determines $\xi$ uniquely and leaves no residual diffeomorphisms. The remaining freedom consists of internal rotations, each combined with the diffeomorphism that restores $a_4=0$; it acts on the remaining field as $\delta a_8=(d^A\omega)_8$. Of the eight components of $a_8$, three are thus removed by gauge freedom and three by the corresponding constraints, leaving the two degrees of freedom of the graviton. Second, in this gauge the metric perturbation is tracefree. Indeed, $\sd^i\wedge d^Aa^i=d(\sd^i\wedge a^i)$ because $d^A\sd^i=0$, and the map $a\mapsto\sd^i\wedge a^i$ annihilates $(\Eoneforms)_8$. Hence the one-dimensional component of $d^Aa$, which by \eqref{eq:lin-ee-curv-cond} carries the trace of the metric perturbation, vanishes when $a_4=0$.

We now describe the effect of the gauge transformations on $\sigma=d_1^*a$ and $\psi=d_2a$. For an internal rotation, $(d^A)^2\omega^i=\epsilon^{ijk}F^j\omega^k=\frac\Lambda3\epsilon^{ijk}\sd^j\omega^k$, so
\begin{align}
	\delta_\omega\sigma^i=\frac3\Lambda(\mathbb I-P_5)\,(d^A)^2\omega^i=\epsilon^{ijk}\sd^j\omega^k ,
	\qquad
	\delta_\omega\psi^{ij}\sd^j=P_5\bigl(\tfrac\Lambda3\epsilon^{ijk}\sd^j\omega^k\bigr)=0 .
\end{align}
The 2-form perturbation changes by an infinitesimal internal rotation of the background triple. This lies in the three-dimensional summand and does not change the metric. For a diffeomorphism, \eqref{eq:gauge-diffeo} gives
\begin{align}
	\delta_\xi\sigma^i=(\mathbb I-P_5)\,d^A(i_\xi\sd^i),\qquad \delta_\xi\psi^{ij}\sd^j=P_5\,d^A(i_\xi\sd^i)=0 .
\end{align}
Since $d^A\sd^i=0$, $d^A(i_\xi\sd^i)$ is the covariant Lie derivative of the background triple: the linearised diffeomorphism of $\sd$, up to an internal rotation. Its metric part is $2\nabla_{(\mu}\xi_{\nu)}$, as verified in Section~\ref{sec:conformal-reconstruction}. The vanishing of its five-dimensional part expresses the fact that diffeomorphisms preserve metricity, \eqref{eq:lin-metricity-cond}. Thus $\sigma$ transforms as the 2-form encoding of the metric should: by linearised diffeomorphisms and internal frame rotations. The curvature $\psi$ is gauge invariant. For diffeomorphisms this is the statement that the linearised Weyl curvature of a perturbation of a conformally flat background is gauge invariant; for internal rotations it holds because $\psi$ transforms in the adjoint representation and the background Weyl curvature vanishes. Correspondingly, both gauge directions are annihilated by $E_a=d_2^*d_2$, as they must be.

\section{Conformal Geometry and Reconstruction}\label{sec:conformal-reconstruction}

The preceding section establishes why potential reconstruction works. We now use the stronger geometric information supplied by conformal flatness to choose an effective reconstruction operator. We first express the background connection through a conformal factor, then compute the potential operator and introduce a weight that removes its component coupling. Finally, we relate the result to the Teukolsky--Wald identity and extract the metric perturbation explicitly.

\subsection{Conformal Structure of Constant-Curvature Backgrounds}

A constant-curvature metric is locally conformally flat~\cite{Introduction_to_Lee_J_2018}. We work on a patch admitting a nonvanishing conformal factor and a flat representative $\hat g$. We fix the convention $\hat g=\lambda^2g$. The transformation of the chiral variables makes it possible to express the connection terms through derivatives of this factor.

The self-dual two-forms have the same conformal weight as the metric:
\begin{align}
	\hat g = \lambda^2 g, \qquad g=\lambda^{-2}\hat g, \qquad \sd^i = \lambda^{-2}\hat{\sd}^i.
\end{align}
The conformal transformation on the connection in general can be computed to find,
\begin{align}
	A^i = \hat{A}^i - \hat{\sd}^i \cdot d \log(\lambda).
\end{align}
The dot denotes contraction of a one-form with the mixed-index endomorphism, $(\hat\sd^i\cdot X)_\mu=\hat\sd^i_\mu{}^\nu X_\nu$. Choose a parallel self-dual frame for the flat representative, so that $\hat A=0$. The connection formula then becomes
\begin{align}\label{eq:max-sym-con}
	A^i = - \sd^i \cdot d \log(\lambda).
\end{align}
Here we used the fact that the mixed-index endomorphisms are conformally invariant, $\sd^i_\mu{}^\nu=\hat\sd^i_\mu{}^\nu$. This is why the background connection can be expressed entirely through the conformal factor and the self-dual frame.

\subsection{Curvature Wave Operator}

We first compute the composition $d_2 d_2^*$ to identify the terms that prevent decoupling. In this calculation $\psi$ denotes an arbitrary symmetric tracefree argument of the operator; when used to generate a solution, it plays the role of the potential $\phi$ in the introduction, and is not assumed to satisfy the first-order curvature equation.

Consider
\begin{align}
	(d_2 d_2^*)(\psi){}^i = -\frac{3\im}{\Lambda} (P_5 d^A J_1^{-1} \star d^A)(\psi \sd){}^i.
\end{align}
Using (\ref{d2-star}), the composition $d_2 d_2^*$ simplifies to become,
\begin{align}
	(d_2 d_2^*)(\psi){}^i = \frac{3}{\Lambda} (P_5 d^A \sd \cdot d^A)(\psi){}^i. 
\end{align}
We now identify the five-dimensional two-form space with ${\rm Sym}^2_0(\C^3)$ using \cref{eq:sym-0-C3-extract}. The quaternion algebra separates the composition into a covariant wave operator and a commutator of covariant derivatives:
\begin{align}
	(d_2 d_2^*)(\psi){}^{ij} & = \frac{3}{\Lambda} \frac{1}{4} \sd^{\langle j |\mu\nu} 2 \nabla^A_\mu (\sd^k_\nu{}^\rho \nabla^A_\rho \psi^{i\rangle k}) = \frac{3}{2\Lambda} \left( -\delta^{\langle j | k} g^{\mu\nu} + \epsilon^{\langle j | k l} \sd^{l\mu\nu} \right) \nabla^A_\mu \nabla^A_\nu \psi^{|i \rangle k} \\
							 & = -\frac{3}{2\Lambda} \Box^A \psi^{ij} + \frac{3}{2\Lambda} \epsilon^{\langle i | k l} \sd^{l\mu\nu} \nabla^A_\mu \nabla^A_\nu \psi^{| j \rangle k}.
\end{align}
From here we can make use of the following,
\begin{align}
	\nabla^A_{[\mu} \nabla^A_{\nu]} \psi^{ij} = \frac{1}{2} \epsilon^{ikl} F^k_{\mu\nu} \psi^{lj} + \frac{1}{2} \epsilon^{jkl} F^k_{\mu\nu} \psi^{il} = \frac{\Lambda}{6}  \epsilon^{ikl} \sd^k_{\mu\nu} \psi^{lj} +\frac{\Lambda}{6} \epsilon^{jkl} \sd^k_{\mu\nu} \psi^{il},
\end{align}
to find that
\begin{align}
	(d_2 d_2^*)(\psi){}^{ij} & = -\frac{3}{2\Lambda} \Box^A \psi^{ij} + \frac{3}{2\Lambda} \frac{\Lambda}{6} \epsilon^{ikl} \sd^{l\mu\nu} ( \epsilon^{jmn} \sd^m_{\mu\nu} \psi^{nk} + \epsilon^{kmn} \sd^m_{\mu\nu} \psi^{jn}) \\
							 & = -\frac{3}{2\Lambda} ( \Box^A \psi^{ij} - 2 \Lambda \psi^{ij} ).
\end{align}
The resulting operator contains the gauge-covariant wave operator $\Box^A=\nabla^{A\mu}\nabla^A_\mu$ and a curvature-dependent zeroth-order term. Although its principal part is scalar, the internal connection mixes the matrix components of the argument. Thus it is not yet a system of independent scalar wave equations. Our objective is to remove this mixing by modifying the reconstruction map.

\subsection{Conformal Wave Operator}

To find the required modification, substitute the conformal expression \cref{eq:max-sym-con} into $d_2$ and $d_2^*$. For the first one we have:
\begin{align}\nonumber
	d_2(a){}^{ij}  = \frac{1}{2} \sd^{\langle j|\mu\nu} \nabla^A_\mu a^{|i \rangle}_\nu =  \frac{1}{2} \sd^{\langle j|\mu\nu} (\nabla_\mu a^{|i\rangle}_\nu+\epsilon^{|i\rangle kl} A^k_\mu a^{l}_\nu)=	\frac{1}{2} \sd^{\langle j|\mu\nu} (\nabla_\mu a^{|i\rangle}_\nu-\epsilon^{|i\rangle kl} \Sigma^k_\mu{}^\rho \nabla_\rho \log(\lambda)  a^{l}_\nu)\\
	=\frac{1}{2} \sd^{\langle j|\mu\nu} (\nabla_\mu + \nabla_\mu \log(\lambda)) a^{|i \rangle}_\nu = \frac{1}{2} \lambda^{-1} \sd^{\langle j | \mu\nu} \nabla_\mu ( \lambda a^{|i\rangle}_\nu ).
\end{align}	
For the second operator we have, using (\ref{d2-star})	
\begin{align}	\nonumber
	d_2^*(\psi){}^i_\mu  = \frac{3}{\Lambda} \sd^j_\mu{}^\nu \nabla^A_\nu \psi^{ij} = \frac{3}{\Lambda} \sd^j_\mu{}^\nu (\nabla_\nu \psi^{ij}+ \epsilon^{ikl} A^k_\nu \psi^{lj} + 
	\epsilon^{jkl} A^k_\nu \psi^{il} ) = \\ \nonumber
	\frac{3}{\Lambda} \sd^j_\mu{}^\nu (\nabla_\nu \psi^{ij}- \epsilon^{ikl} \Sigma^k_\nu{}^\rho \nabla_\rho \log(\lambda) \psi^{lj} - 
	\epsilon^{jkl} \Sigma^k_\nu{}^\rho \nabla_\rho \log(\lambda) \psi^{il} ) 
	\\
	=\frac{3}{\Lambda} \sd^j_\mu{}^\nu (\nabla_\nu - 3 \nabla_\nu \log(\lambda) ) \psi^{ij} = \frac{3}{\Lambda} \lambda^3 \sd^j_\mu{}^\nu \nabla_\nu (\lambda^{-3} \psi^{ij}). 
\end{align}
These weighted derivatives suggest inserting a scalar multiplication between $d_2$ and $d_2^*$ so that the intermediate weights cancel. To see the required factor, let us rewrite both operators explicitly using the flat objects
\begin{align}
d_2(a){}^{ij}  = \frac{1}{2} \lambda \hat{\sd}^{\langle j | \mu\nu}\hat{\nabla}_\mu ( \lambda a^{|i\rangle}_\nu ), \qquad
d_2^*(\psi){}^i_\mu  =  \frac{3}{\Lambda} \lambda^3 \hat{\sd}^j_\mu{}^\nu \hat{\nabla}_\nu (\lambda^{-3} \psi^{ij}).
\end{align}
To write the first one we used $\Sigma^{i\mu\nu} = \lambda^2 \hat{\Sigma}^{i\mu\nu}$, to write the second we used the conformal invariance of the mixed object $\Sigma^i_\mu{}^\nu = \hat{\Sigma}^i_\mu{}^\nu$. These expressions make it clear that inserting the factor of $\lambda^{-4}$ between $d_2$ and $d_2^*$ cancels the factors of $\lambda$ between the two derivatives, which produces the flat box operator
\begin{align}
	(d_2 \lambda^{-4} d_2^*)(\psi){}^{ij} & = \frac{3}{2\Lambda} \lambda \hat{\sd}^{\langle j | \mu\nu} \hat{\sd}^k_\nu{}^\rho \hat{\nabla}_\mu \hat{\nabla}_\rho (\lambda^{-3} \psi^{|i\rangle k}) = \frac{3}{2 \Lambda} \lambda (-\delta^{\langle j | k} \hat{g}^{\mu\nu} + \epsilon^{\langle j | kl} \hat{\sd}^{l\mu\nu}) \hat{\nabla}_\mu \hat{\nabla}_\nu (\lambda^{-3} \psi^{|i\rangle k}) \\
										 & = -\frac{3}{2\Lambda}\lambda \,\hat{\Box}(\lambda^{-3} \psi^{ij}).
\end{align}
The calculation reduces the weighted equation to a wave equation for a rescaled matrix of potentials. Because the flat box operator decouples, its components can then be solved independently.

The weighted composition thus determines the potential operator and the reconstruction map:
\begin{align}
	\mathcal{P} = d_2 \circ \lambda^{-4} \circ d_2^* \quad \Rightarrow \quad d_3 = \lambda^{-4} d_2^*.
\end{align}
Writing the potential as $\phi^{ij}=\lambda^3\chi^{ij}$ makes the scalar equation and reconstruction explicit:
\begin{align}\label{P-flat-Box}
 \mathcal P(\lambda^3\chi)^{ij}&=-\frac3{2\Lambda}\lambda \hat\Box\chi^{ij},\\
 a_\mu^i= d_3(\lambda^3\chi)^i_\mu&=\frac3\Lambda\lambda^{-1} \sd^j_\mu{}^\nu\partial_\nu\chi^{ij}.
\end{align}
Thus each component of the rescaled potential $\chi$ satisfies the flat wave equation. Here the derivative acts on scalar components in the parallel self-dual frame of $\hat g$.

\subsection{Relation to the Teukolsky--Wald Identity}

We now explain how the second-order Teukolsky--Wald relation follows from the first-order identity. In this subsection $\mathcal O$ denotes the curvature-decoupling operator in the standard Teukolsky--Wald notation,
\begin{align}\label{eq:teuk-wald-iden}
	\mathcal{S} \mathcal{G} = \mathcal{O} \mathcal{T}.
\end{align}
Here $\mathcal G$ acts on 2-form perturbations, $\mathcal T$ extracts their chiral curvature, and $\mathcal S$ maps the Einstein-equation space to the curvature space. Taking the formal adjoint gives the reconstruction identity $\mathcal G\mathcal S^*=\mathcal T^*\mathcal O^*$, with the potential equation $\mathcal P\phi=0$ and $\mathcal P=\mathcal O^*$. For a scalar weight $w=\lambda^{-4}$, the operator $d_2wd_2^*$ is formally self-adjoint under the bilinear pairings used here. Thus the same weighted expression appears in both roles in this case. The assignments below exhibit the factorisation on the 2-form space.
\begin{align}
	\mathcal{G} & = d_1^* d_1 - \mathbb{I}, \\
	\mathcal{T} & = d_2 d_1, \\
	\mathcal{O} & = d_2 \lambda^{-4} d_2^*, \\ 
	\mathcal{S} & = d_2 \lambda^{-4} d_1.
\end{align}
Looking at the right-hand side of \cref{eq:teuk-wald-iden} we find
\begin{align}
	\mathcal{O} \mathcal{T} = d_2 \lambda^{-4} d_2^* d_2 d_1,
\end{align}
then using the pure connection version of the TW identity \cref{eq:pure-con-wald-identity} we find
\begin{align}
	\mathcal{O} \mathcal{T} = d_2 \lambda^{-4} (d_1 d_1^* - \mathbb{I}) d_1 = (d_2 \lambda^{-4} d_1) (d_1^* d_1 - \mathbb{I}) = \mathcal{S} \mathcal{G}.
\end{align}

The second-order identity is therefore a consequence of \cref{eq:pure-con-wald-identity}. In the reverse direction, the reconstruction map factors as $\mathcal S^*=d_1^*\lambda^{-4}d_2^*=d_1^*d_3$. It can be read as first generating the connection, and then the 2-form perturbation. The difference with the standard metric description is that these operators act on constrained two-forms rather than directly on symmetric tensors. 

\subsection{Explicit Metric Reconstruction}

We now translate the above prescription for the 2-form reconstruction into an explicit formula for the symmetric metric perturbation. We will see that the metric perturbation reconstructed via our scheme is tracefree. We will then comment on the natural possibility of adding the trace, by using the gauge freedom.

The metric perturbation is reconstructed from the $\sigma$ perturbation as follows. Let $e^I$ be a background tetrad and $h^I=\delta e^I$ its variation. In the formulas immediately below, $h_{\mu\nu}:=h^I_\mu e_{I\nu}$ denotes the tetrad variation with its internal index converted to a spacetime index; it need not be symmetric. The metric variation is
\begin{align}
	\delta g_{\mu\nu} = 2 h_{(\mu\nu)} = 2 h^I_{(\mu} e_{|I|\nu)}.
\end{align}
The 2-form perturbation $\sigma^i$ can be written with this frame as,
\begin{align}
	\sigma^i_{\mu\nu} = 2 \sd^i_{[\mu}{}^\rho h_{\nu]\rho}.
\end{align}
The sixteen-component tetrad variation contains frame rotations as well as the metric variation. To extract the latter directly from the chiral two-form perturbation, introduce the algebraic map $P_{10}:E\otimes\Lambda^2\to\operatorname{Sym}^2(T^*M)$,
\begin{align}
	P_{10}(B)_{\mu\nu} = \sd^i_{(\mu}{}^\rho B^i_{\nu)\rho} - \frac{1}{6} g_{\mu\nu} \sd^{i\rho\sigma} B^i_{\rho\sigma}.
\end{align}
Using this operator we find that $P_{10}(\sigma){}_{\mu\nu} = \delta g_{\mu\nu}$ provides an explicit map that recovers the metric part of the perturbation.

Now we can extract the metric perturbation from the full 2-form recovery operator $d_1^*$. In doing so we note the identity,
\begin{align}
	P_{10} (\mathbb{I} - P_5) = P_{10}.
\end{align}
The Weyl-type component is annihilated by $P_{10}$, so the preceding projection in $d_1^*$ can be omitted when computing the metric. Consequently,
\begin{align}
	\delta g_{\mu\nu} = (P_{10} d_1^*)(a){}_{\mu\nu} = \frac{3}{\Lambda} P_{10}(d^A a){}_{\mu\nu}.
\end{align}
In components this becomes,
\begin{align}
	\delta g_{\mu\nu} = \frac{3}{\Lambda} \sd^i_{(\mu}{}^\rho ( \nabla^A_{\nu)} a^i_\rho - \nabla^A_{|\rho|} a^i_{\nu)} ) - \frac{1}{\Lambda} g_{\mu\nu} \sd^{i\rho\sigma} \nabla^A_\rho a^i_\sigma.
\end{align}

For reconstruction from a potential, substitute $a=d_3\phi$, where $\phi$ is a symmetric tracefree potential, not the extracted curvature of $a$. The component form of the weighted adjoint is
\begin{align}
	a^i_\mu = d_3(\phi){}^i_\mu = -\frac{3\im}{\Lambda} \lambda^{-4} (J_1^{-1} \star d^A)(\phi\sd){}^i_\mu = \frac{3}{\Lambda} \lambda^{-4} \sd^j_\mu{}^\nu \nabla^A_\nu \phi^{ij}.
\end{align}
Composing this expression with $P_{10}d_1^*$ gives the second-order metric reconstruction operator. 

To compute the metric reconstruction operator explicitly, we use the fact noted in (\ref{a-8}) that the connection generated from a symmetric tracefree argument belongs to the eight-dimensional summand. This has the following useful consequence
\begin{align}
	\sd^i_\mu{}^\nu \nabla^A_\rho a^i_\nu = 0.
\end{align}
This lets us simplify the metric reconstruction from the potential considerably, indeed we have
\begin{align}
	\delta g_{\mu\nu} = - \frac{3}{\Lambda} \sd^i_{(\mu}{}^\rho \nabla^A_{|\rho|} a^i_{\nu)} = - \left( \frac{3}{\Lambda} \right)^2 \sd^i_{(\mu}{}^\rho \sd^j_{\nu)}{}^\sigma \nabla^A_\rho ( \lambda^{-4} \nabla^A_\sigma \phi^{ij}).
\end{align}
Passing the self-dual 2-forms through the covariant derivatives allows us to write this in a more familiar form,
\begin{align}
	\delta g_{\mu\nu} = - \left( \frac{3}{\Lambda} \right)^2 \nabla^\rho (\lambda^{-4} \nabla^\sigma \mathcal{H}_{(\mu|\rho |\nu )\sigma} ), \quad \mathcal{H}_{\mu\rho\nu\sigma} = \phi^{ij} \sd^i_{\mu\rho} \sd^j_{\nu\sigma}
\end{align}
The explicit symmetrisation is needed: the weighted double divergence of $\mathcal H$ is not by itself symmetric in $\mu\nu$.
This double-divergence form makes contact with the Bergmann--Sachs type of metric reconstruction~\cite{Structure_of_Pa_Sachs_1958,Linearized_grav_Akstei_2010}. For the tracefree potential considered so far, the resulting metric perturbation is tracefree:
\begin{align}
	g^{\mu\nu} \delta g_{\mu\nu} = 0.
\end{align}

To include a trace in the reconstructed metric, which is important in applications, we first observe that the trace of a general reconstructed perturbation depends on the contraction selecting the four-dimensional part of the connection:
\begin{align}
	g^{\mu\nu} \delta g_{\mu\nu} = \frac{2}{\Lambda} \sd^{i\mu\nu} \nabla^A_\mu a^i_\nu.
\end{align}
On the constant curvature background, the four-dimensional connection summand describes infinitesimal diffeomorphism directions, up to internal gauge transformations. Its elements have the form $\sd^i\cdot\xi$ for a one-form $\xi$. A scalar trace added to the potential generates a particular family of such directions. Replace the tracefree potential $\phi^{ij}$ by $\phi^{ij}+\delta^{ij}f$, where $f$ is an additional scalar. Extending the same differential expression gives
\begin{align}
	a^i_\mu = d_3(\phi+\delta f){}^i_\mu = \frac{3}{\Lambda} \lambda^{-4} \sd^j_\mu{}^\nu \nabla^A_\nu (\phi^{ij}+\delta^{ij}f).
\end{align}
The trace part does not require an additional potential equation: its contribution is annihilated by $d_2d_3$, and hence by the connection Einstein operator. Explicitly,
\begin{align}
 (d_2d_3)(\delta f)^{ij}
 &=\frac{3}{2\Lambda}\sd^{\langle j|\mu\nu}\nabla^A_\mu
   \bigl(\lambda^{-4}\sd^{|i\rangle}_\nu{}^\rho\nabla_\rho f\bigr)\\
 &=-\frac{3}{2\Lambda}\delta^{\langle ij\rangle}
    \nabla^\mu(\lambda^{-4}\nabla_\mu f)=0.
\end{align}
The tracefree simplification used above no longer applies to this additional connection component. Instead, applying the full metric projection gives
\begin{align}
 \delta g_{\mu\nu}[\phi+\delta f]
 &= -\left(\frac3\Lambda\right)^2\nabla^\rho
       \bigl(\lambda^{-4}\nabla^\sigma
       (\phi^{ij}\sd^i_{(\mu|\rho|}\sd^j_{\nu)\sigma})\bigr)
    -2\left(\frac3\Lambda\right)^2\nabla_{(\mu}
       (\lambda^{-4}\nabla_{\nu)}f).
\end{align}
The second term is the diffeomorphism perturbation $2\nabla_{(\mu}\xi_{\nu)}$ with $\xi_\mu=-(3/\Lambda)^2\lambda^{-4}\nabla_\mu f$. Thus the scalar trace changes a gauge part of the reconstructed field, without parametrising every possible gauge transformation. More generally, (see Section~\ref{subsec:gauge}), every connection perturbation of the form $a^i=\sd^i\cdot\zeta$, for an arbitrary one-form $\zeta$, satisfies $d_2a=0$, hence $E_aa=0$, and generates the linearised diffeomorphism $\delta g_{\mu\nu}=2\nabla_{(\mu}\xi_{\nu)}$ with $\xi=-(3/\Lambda)\zeta$. Similarly, internal gauge transformations $a^i=(d^A\omega)^i$ satisfy $d_2a=P_5(\epsilon F\omega)=0$ and do not change the metric. The general gauge transformation can therefore be added directly at the level of the connection. This may be useful for metric reconstruction in a specific gauge. 

\subsection{Example: Single-Helicity Perturbations of de Sitter Space}\label{sec:example-ds}

This subsection illustrates the conformal reconstruction in explicit coordinates and compares it with the scalar ansatz of~\cite{Pure_connection_Krasno_2024}. The calculation starts from the conformally flat metric
\begin{align}
	g = \frac{1}{t^2} (-dt^2 + dx^i dx^i)
\end{align}
in a patch coordinatised by $(t,x^1=x,x^2=y,x^3=z)$. The background self-dual 2-forms are given by
\begin{align}\label{ds-sigma}
	\sd^i = \frac{1}{t^2} ( \im dt \wedge dx^i - \frac{1}{2} \epsilon^{ijk} dx^j \wedge dx^k).
\end{align}
The background connections for these 2-forms are
\begin{align}
	A^i_\mu = -\sd^i_\mu{}^\nu \partial_\nu \log(t).
\end{align}
From this we read off $\lambda = t$; indeed, $t^2 g = \eta$ is the flat Minkowski metric. The curvatures are
\begin{align}
	F^i = \sd^i,
\end{align}
Thus $\Lambda=3$ in the convention $F^i=(\Lambda/3)\sd^i$. The displayed Lorentzian metric is the unit-radius de Sitter metric. Since $\hat g=t^2g$ is Minkowski, its wave operator is
\begin{align}
	\hat{\Box} = -\partial_t^2 + \partial_i \partial_i.
\end{align}

Choose a symmetric tracefree matrix $\chi^{ij}$ of scalar solutions of this flat wave equation. The weighted first-order reconstruction then gives the connection in the following form:
\begin{align}
	a^i = d_3(\lambda^3 \chi) =  \lambda^{-4} d_2^*(\lambda^3 \chi) = \lambda^{-1} \sd^j \cdot d \chi^{ij} = t^{-1} ( \im dt \partial^j \chi^{ij} + \im dx^j \partial_t \chi^{ij} - \epsilon^{jlm} dx^l \partial^m \chi^{ij}),
\end{align}
where we used $\Lambda=3$. To compare with~\cite{Pure_connection_Krasno_2024}, take the rank-one potential $\chi^{ij}=S^iS^jf$ with $S^i=(0,1,\im)$. Since $S^iS^i=0$, this matrix is tracefree. The resulting connection has $a^1=0$ and $a^3=\im a^2$, so $a^-=a^2-\im a^3=2a^2$. Explicitly,
\begin{align}
	a^- = 2\im t^{-1} (\partial_y f + \im \partial_z f) (dt - dx) + 2\im t^{-1} (\partial_t f + \partial_x f)(dy + \im dz).
\end{align}
After the corresponding Wick rotation, coordinate change, and an overall normalisation of the scalar potential, this reproduces the ansatz of~\cite{Pure_connection_Krasno_2024}. The general matrix potential allows a wider class of inputs, while the rank-one choice displays the relation to the scalar potentials appearing in the linearisation of \pleb{}'s heavenly equations.

Applying $P_{10}d_1^*$ to the connection above gives the metric perturbation explicitly. Introduce the null one-forms $l=dt-dx$ and $m=dy+\im dz$ of the flat metric $\hat g$, and the derivatives $\partial_n=\partial_t+\partial_x$ and $\partial_m=\partial_y+\im\partial_z$; up to sign, $\partial_n$ and $\partial_m$ are the vectors $\hat g$-dual to $l$ and $m$. For the rank-one potential $\chi^{ij}=S^iS^jf$, and using $\hat\Box f=0$, one finds
\begin{equation}\label{eq:ds-metric}
\delta g=\frac1t\,(\partial_m^2 f)\,l\otimes l+2\,\partial_n\Big(\frac{\partial_m f}{t}\Big)\,l\odot m+t\,\partial_n\Big(\frac{\partial_n f}{t^2}\Big)\,m\otimes m,
\qquad l\odot m=\tfrac12(l\otimes m+m\otimes l).
\end{equation}
We have checked by direct computation that \eqref{eq:ds-metric} satisfies the linearised Einstein equations $\delta(R_{\mu\nu}-\Lambda g_{\mu\nu})=0$, with $\Lambda=3$, for solutions of the flat wave equation $\hat\Box f=0$. The metric description of self-dual perturbations of constant-curvature backgrounds has been obtained before in the context of \pleb{}'s second heavenly equation~\cite{Lipstein:2023sdads,Pure_connection_Krasno_2024}. We include \eqref{eq:ds-metric} because it shows directly how the first-order construction produces a single-helicity, and therefore complex, metric perturbation.

Three features of \eqref{eq:ds-metric} are worth noting. First, $\delta g$ is tracefree and takes values in the symmetric square of the span of $l$ and $m$. These one-forms are null and mutually orthogonal, so they span a totally null two-plane, and the derivatives of $f$ that appear are derivatives along the same plane. This is the structure familiar from \pleb{}'s second heavenly equation, where the metric perturbation is built from second derivatives of a single scalar along a totally null plane. Second, in Lorentzian signature a totally null two-plane is necessarily complex: the null one-form $m$ cannot be chosen real. The single-helicity perturbation generated by $\mathcal P\phi=0$ is therefore complex for any choice of $f$, including a real one. This is the concrete form of the reality issue discussed in Section~\ref{sec:discussion}. In split signature totally null planes can be real, and the analogous perturbations are real. Third, the real part $\mathrm{Re}\,\delta g$ is a real solution of the linearised Einstein equations, but it combines both helicities: its self-dual Weyl curvature is the complex conjugate of the anti-self-dual Weyl curvature of $\delta g$ (up to a factor), and is therefore nonzero. General potentials $\chi^{ij}$ superpose contributions of this type associated with different null vectors $S^i$ and different totally null planes.

\section{First-Order Equations with Sources}\label{sec:sourced-equations}

Having established the vacuum equations and their operator identities, we can now extend the construction to nonzero stress-energy. We continue to work on a constant-curvature vacuum background with $\Lambda\neq0$. The operators and their formal adjoints remain those of Sections~\ref{sec:linear-eom} and~\ref{sec:conformal-reconstruction}; matter changes their sources and the reconstruction formula, rather than their definitions. The 2-form and connection equations remain equivalent. Curvature reconstruction requires a further step: the potential equation is now inhomogeneous, with the actual Weyl curvature as its source. We work out the algebraic details of sourced reconstruction here. Subtler issues of reality and comparison with the usual schemes are postponed to Section~\ref{sec:discussion}. 

\subsection{Stress-Energy as a Curvature Source}

Let $\tau_{\mu\nu}$ be the first-order stress-energy tensor, normalised by
\begin{align}\label{eq:sourced-metric-einstein}
 \delta(G_{\mu\nu}+\Lambda g_{\mu\nu})=\kappa \tau_{\mu\nu},
 \qquad \kappa=8\pi G.
\end{align}
All contractions and covariant derivatives in this section use the background metric. We assume a symmetric stress-energy tensor and no independent spin current, so the compatibility equation remains torsion-free. To express the matter contribution as an $E$-valued two-form, define the algebraic map
\begin{align}\label{eq:source-U-map}
 U(q)^i_{\mu\nu}:=q_{[\mu}{}^\rho\sd^i_{|\rho|\nu]},
 \qquad q_{\mu\nu}=q_{\nu\mu}.
\end{align}
It obeys $U(g)=\sd$ and $P_{10}U(q)=q$. The trace of $q$ maps to the one-dimensional summand of $E\otimes\Lambda^2$, and its tracefree part maps to the nine-dimensional anti-self-dual summand. In particular, $P_5U(q)=0$.

Write $\tau=g^{\mu\nu}\tau_{\mu\nu}$ and $\tau^{\mathrm{TF}}_{\mu\nu}=\tau_{\mu\nu}-\frac14 \tau g_{\mu\nu}$. The Ricci decomposition of the curvature gives the source two-form
\begin{align}\label{eq:matter-curvature-source}
 \tau^i
 &=\kappa U(\tau^{\mathrm{TF}})^i-\frac{\kappa \tau}{12}\sd^i
 =\kappa U(q)^i,\qquad q_{\mu\nu}=\tau_{\mu\nu}-\frac13 \tau g_{\mu\nu},
 \qquad P_5\tau^i=0.
\end{align}
The scalar coefficient follows from $\delta R=-\kappa \tau$, since the scalar-curvature contribution to the chiral curvature is $(R/12)\sd^i$. The tracefree Ricci contribution is $U(\operatorname{Ric}^{\mathrm{TF}})^i$. These observations fix both the relative trace coefficient and the normalisation in \cref{eq:matter-curvature-source}.

The linearised equations are therefore
\begin{align}\label{eq:sourced-plebanski}
 P_5\sigma&=0,\qquad
 d^A\sigma^i+\epsilon^{ijk}a^j\wedge\sd^k=0,\\
 d^Aa^i&=\psi^{ij}\sd^j+\frac\Lambda3\sigma^i+\tau^i.
 \label{eq:sourced-curvature-decomposition}
\end{align}
Because the background stress-energy vanishes, the source two-form is evaluated using the background triple; varying that triple in the source would contribute only at higher perturbative order.

\subsection{Equivalent Sourced 2-Form and Connection Equations}

It is convenient to absorb the factor used in $d_1^*$ into a normalised 2-form source,
\begin{align}\label{eq:normalised-frame-source}
 j^i:=\frac3\Lambda\tau^i.
\end{align}
Compatibility still gives $a=d_1\sigma$. Projecting \cref{eq:sourced-curvature-decomposition} onto the five-dimensional summand and its complement gives, respectively,
\begin{align}\label{eq:sourced-first-order-system}
 \psi=d_2a,\qquad \sigma=d_1^*a-j,\qquad a=d_1\sigma.
\end{align}
Here and below, as in the vacuum discussion, $\psi^{ij}$ is identified with the two-form $\psi^{ij}\sd^j$ when used as an argument of the operators. The algebraic source correction in the reverse map is essential.

Eliminating the connection or the 2-form perturbation yields
\begin{align}\label{eq:sourced-frame-connection}
 E_\sigma\sigma=j,\qquad
 E_a a=d_1j,\qquad
 E_\sigma=d_1^*d_1-\mathbb I,\qquad
 E_a=d_1d_1^*-\mathbb I.
\end{align}
These two descriptions are equivalent, including their source terms. Indeed, if $a$ solves the connection equation and $\sigma=d_1^*a-j$, then
\begin{align}
 d_1\sigma=d_1d_1^*a-d_1j=a,
 \qquad E_\sigma\sigma=d_1^*a-\sigma=j.
\end{align}
Conversely, a 2-form solution gives $a=d_1\sigma$, for which
\begin{align}
 E_a a=d_1E_\sigma\sigma=d_1j,
 \qquad d_1^*a-j=\sigma.
\end{align}
Thus the inverse correspondence on vacuum solution spaces becomes an affine correspondence for a fixed source.

It is illuminating to write the formula for the metric perturbation as determined by a connection and source explicitly. Applying the metric projection gives
\begin{align}\label{eq:sourced-metric-reconstruction}
 \delta g_{\mu\nu}
 &=(P_{10}d_1^*a)_{\mu\nu}-(P_{10}j)_{\mu\nu}\nonumber\\
 &=(P_{10}d_1^*a)_{\mu\nu}
 -\frac{3\kappa}{\Lambda}\left(\tau_{\mu\nu}-\frac13\tau g_{\mu\nu}\right).
\end{align}
For a distributional stress-energy tensor this equation is interpreted distributionally. 

\subsection{The Sourced First-Order Curvature Equation}

The operator identity \cref{eq:pure-con-wald-identity} is a background identity and is unchanged by the source. Combining it with \cref{eq:sourced-frame-connection} and $\psi=d_2a$ gives
\begin{align}\label{eq:sourced-first-order-bianchi}
 d_2^*\psi=d_1j.
\end{align}
This is a first-order equation for the actual chiral Weyl curvature. Substituting the definitions of $d_2^*$, $d_1$, and $j$ shows that it is equivalent to
\begin{align}\label{eq:sourced-exterior-bianchi}
 d^A(\psi^{ij}\sd^j+\tau^i)=0.
\end{align}
It can also be obtained by differentiating \cref{eq:sourced-curvature-decomposition}: the term $(d^A)^2a$ cancels $(\Lambda/3)d^A\sigma$ by compatibility and $F^i=(\Lambda/3)\sd^i$. Thus \cref{eq:sourced-first-order-bianchi} is the chiral form of the sourced Bianchi equation.

The source must satisfy the conservation law
\begin{align}\label{eq:linear-source-conservation}
 \nabla^\mu \tau_{\mu\nu}=0.
\end{align}
This is required by the linearised Einstein equation and is also a compatibility condition of the sourced curvature system. Applying $d_2$ to \cref{eq:sourced-first-order-bianchi} gives the second-order consequence
\begin{align}\label{eq:sourced-unweighted-curvature-wave}
 d_2d_2^*\psi=d_2d_1j.
\end{align}
As before, there is no subtraction of $\mathbb I$ in this curvature operator. Consequently, the equation does not provide an algebraic inverse to curvature extraction: $d_2^*\psi$ returns the differentiated source, rather than the connection. Moreover, the second-order equation alone is weaker than \cref{eq:sourced-first-order-bianchi}; its solutions must obey the first-order constraints as well.

\subsection{Conformal Reduction of the Sourced Curvature Equation}

The conformal choice from Section~\ref{sec:conformal-reconstruction} remains useful. Set
\begin{align}
 w=\lambda^{-4},\qquad d_3=wd_2^*,\qquad
 \mathcal P=d_2wd_2^*.
\end{align}
Multiplying \cref{eq:sourced-first-order-bianchi} by $w$ and applying $d_2$ yields
\begin{align}\label{eq:sourced-weighted-curvature-wave}
 \mathcal P\psi=d_2wd_1j.
\end{align}
This is the sourced Teukolsky--Wald relation for the operators already constructed: with $\mathcal G=E_\sigma$, $\mathcal T=d_2d_1$, $\mathcal S=d_2wd_1$, and $\mathcal O=\mathcal P$, one has $\mathcal O\psi=\mathcal S j$. The equality $\mathcal O=\mathcal P$ here uses the formal self-adjointness of $d_2wd_2^*$ under the bilinear pairings used here.

In a parallel self-dual frame for $\hat g=\lambda^2g$, the formula (\ref{P-flat-Box}) gives
\begin{align}\label{eq:sourced-flat-curvature-wave}
 \hat\Box(\lambda^{-3}\psi^{ij})
 =-\frac{2\Lambda}{3}\lambda^{-1}(d_2wd_1j)^{ij}.
\end{align}
The left-hand side is the flat wave operator acting separately on the five scalar components. The right-hand side is a second-order differential expression in the stress-energy tensor. This simplification concerns the wave operator, not the elimination of the Bianchi constraints: a solution obtained from \cref{eq:sourced-flat-curvature-wave} must still satisfy \cref{eq:sourced-first-order-bianchi}. However, we expect this constraint to propagate, i.e.\ to hold everywhere if it is imposed on initial data and the source is conserved, as is the case for the massless spin-2 equation in flat space. This suggests that the most efficient way of dealing with the first-order Bianchi constraint is to impose it on initial data. The example of a point mass in de Sitter space in the next section is consistent with this expectation: there the retarded solution satisfies the constraint, including at the worldline. We do not try to prove propagation in general in this paper.

\subsection{Inhomogeneous Potential Reconstruction}\label{subsec:sourced-potential-reconstruction}

The potential reconstruction now differs from the homogeneous construction. Suppose an actual curvature $\psi$ satisfying \cref{eq:sourced-first-order-bianchi} has been found. Instead of imposing $\mathcal P\phi=0$, solve
\begin{align}\label{eq:inhomogeneous-potential-equation}
 \mathcal P\phi=\psi.
\end{align}
On a patch where this equation can be solved with the chosen data, define
\begin{align}\label{eq:sourced-potential-fields}
 a=d_3\phi,\qquad \sigma=d_1^*d_3\phi-j.
\end{align}
Then the extracted curvature is precisely the prescribed one,
\begin{align}
 d_2a=d_2d_3\phi=\mathcal P\phi=\psi,
\end{align}
and the connection satisfies the required sourced equation,
\begin{align}
 E_a a=d_2^*d_2d_3\phi=d_2^*\psi=d_1j.
\end{align}
The 2-form--connection equivalence therefore proves that $\sigma$ solves $E_\sigma\sigma=j$, and \cref{eq:sourced-metric-reconstruction} gives its metric perturbation. The sequence of operations is
\begin{align}\label{eq:sourced-reconstruction-sequence}
 d_2^*\psi=d_1j
 \quad\longrightarrow\quad \mathcal P\phi=\psi
 \quad\longrightarrow\quad a=d_3\phi
 \quad\longrightarrow\quad \sigma=d_1^*a-j.
\end{align}
The distinction between actual curvature and potential is explicit: $\psi$ satisfies the sourced Bianchi equation, while $\phi$ solves a further inhomogeneous equation whose source is $\psi$. Even in a source-free region, a nonzero actual curvature requires this inhomogeneous potential equation. By contrast, $\mathcal P\phi=0$ generates the vanishing-chiral-curvature sector described earlier.

Writing $\phi=\lambda^3\chi$, the additional solve and the connection reconstruction become
\begin{align}\label{eq:sourced-flat-potential-reconstruction}
 \hat\Box\chi^{ij}&=-\frac{2\Lambda}{3}\lambda^{-1}\psi^{ij},\\
 a^i_\mu&=\frac3\Lambda\lambda^{-1} \sd^j_\mu{}^\nu\partial_\nu\chi^{ij}.
\end{align}
Thus the inhomogeneous reconstruction still uses componentwise flat wave equations followed by first-order maps.

For fixed $j$ and $\psi$, one may add any $a_0$ with $d_2a_0=0$, together with the corresponding 2-form perturbation $d_1^*a_0$. Such a field solves $E_a a_0=0$ and changes neither the source nor the extracted curvature. This freedom in our reconstruction mechanism should be kept in mind when dealing with the initial or boundary conditions, gauge freedom, and Lorentzian reality conditions, as we discuss further in the last section. 

\section{Example: a Static Mass in de Sitter Space}\label{sec:example-static-mass}

We illustrate the sourced reconstruction~\eqref{eq:sourced-reconstruction-sequence} with the simplest physical source: a point mass at rest at the origin of the de Sitter patch of Section~\ref{sec:example-ds}. The worldline $x^i=0$ is a geodesic, so the source is conserved. We follow the three steps of~\eqref{eq:sourced-reconstruction-sequence}: we first determine the curvature from the source, then the potential, and then the metric. Only afterwards do we compare the result with the linearised Schwarzschild--de Sitter solution, which serves as an independent check.

\subsection{The Source and the Curvature}\label{subsec:sds-source}

We work in the coordinates of Section~\ref{sec:example-ds}, with $g=t^{-2}(-dt^2+dx^idx^i)$, $t>0$, $\Lambda=3$ and $\lambda=t$, and write
\[
\rho=|x|,\qquad n^i=x^i/\rho, \qquad Y^{ij}=3n^in^j-\delta^{ij}.
\]
The background self-dual 2-forms are given by~(\ref{ds-sigma}), $\bar\sd^i$ is their complex conjugate, and the background connection $A^i=-\sd^i\cdot d\log t$ of~\eqref{eq:max-sym-con} is
\begin{equation}
A^i=-\frac{\im}{t}\,dx^i .
\end{equation}

\paragraph{The source.} A point particle of mass $M$ at rest at $x^i=0$ has 
\[
\tau^{\mu\nu}=M\int u^\mu u^\nu\,\delta^{(4)}(x-z(s))\,(-g)^{-1/2}\,ds,
\]
with $u=t\,\partial_t$ and $ds=dt/t$. The only nonvanishing component of the stress-energy tensor is
\begin{equation}\label{eq:sds-stress}
\tau_{tt}=M\,t\,\delta^{(3)}(x).
\end{equation}
This is conserved, $\nabla^\mu\tau_{\mu\nu}=0$, as it must be for a particle on a geodesic. We write $m=GM$, so that $\kappa\,\tau_{tt}=8\pi m\,t\,\delta^{(3)}(x)$. The tensor $q_{\mu\nu}=\tau_{\mu\nu}-\frac13\tau\,g_{\mu\nu}$ of~\eqref{eq:matter-curvature-source} is then 
\[
\kappa q=\frac{8\pi m}{3}\,t\,\delta^{(3)}(x)\big(2\,dt^2+dx^idx^i\big), 
\]
and with $U(dt\otimes dt)^i=-\frac\im2\,dt\wedge dx^i$ (see Section~\ref{subsec:sds-comparison}) the source 2-form $j^i=\frac3\Lambda\tau^i=\kappa U(q)^i$ is
\begin{equation}\label{eq:sds-j}
j^i=-\frac{4\pi m}{3}\,t\,\delta^{(3)}(x)\Big(\im\,dt\wedge dx^i+\epsilon^{ijk}dx^j\wedge dx^k\Big).
\end{equation}

\paragraph{The curvature.} The curvature obeys the conformally reduced wave equation~\eqref{eq:sourced-flat-curvature-wave}. Evaluating its right-hand side for the source~\eqref{eq:sds-j} gives
\begin{equation}\label{eq:sds-curvature-wave}
\hat\Box\big(t^{-3}\psi^{ij}\big)=-2t^{-1}\big(d_2wd_1j\big)^{ij}
=-4\pi m\Big(\partial_i\partial_j-\frac13\delta^{ij}\nabla^2\Big)\delta^{(3)}(x),
\end{equation}
where $\nabla^2$ is the flat Laplacian. Two features of~\eqref{eq:sds-curvature-wave} determine its solution.

First, the right-hand side is independent of $t$: after the conformal rescaling the source is static. We therefore look for a static rescaled curvature $\hat\psi:=t^{-3}\psi$. For a $t$-independent field $\hat\Box=-\partial_t^2+\nabla^2$ reduces to $\nabla^2$, and~\eqref{eq:sds-curvature-wave} becomes the Poisson equation
\[
\nabla^2\hat\psi^{ij}=-4\pi m\,D_{ij}\,\delta^{(3)}(x),\qquad D_{ij}:=\partial_i\partial_j-\frac13\delta_{ij}\nabla^2 .
\]
This static solution is also the retarded one. The retarded Green's function of $\hat\Box$, applied to a source that is time independent and present for all past times on the patch, reduces to the Green's function of the Laplacian, exactly as for the Coulomb field of a static charge.

Second, the derivatives of $\delta^{(3)}$ on the right-hand side cause no difficulty. The operator $D_{ij}$ has constant coefficients and commutes with $\nabla^2$, so a solution is obtained by applying $D_{ij}$ to the Green's function, rather than by differentiating the delta function: since $\nabla^2\rho^{-1}=-4\pi\delta^{(3)}(x)$, the field $\hat\psi=m\,D_{ij}\rho^{-1}$ solves the Poisson equation in the sense of distributions. This gives
\begin{equation}\label{eq:sds-psi}
\psi^{ij}=m\,t^3\Big(\partial_i\partial_j-\frac13\delta^{ij}\nabla^2\Big)\frac1\rho=m\,t^3\,\frac{Y^{ij}}{\rho^3}.
\end{equation}
In the last expression, $Y^{ij}/\rho^3$ is understood as a principal value, which is well defined because the angular average of $Y^{ij}$ vanishes. The distributional term of $\partial_i\partial_j\rho^{-1}$, namely $-\frac{4\pi}3\delta^{ij}\delta^{(3)}(x)$, is cancelled by the trace subtraction, so that $\psi$ has no contact term at the origin. Thus the second derivatives of $\delta^{(3)}$ in~\eqref{eq:sds-curvature-wave} are completely accounted for by the $\rho^{-3}$ singularity of the Coulomb field, and no derivative of $\delta^{(3)}$, nor $\delta^{(3)}$ itself, appears in $\psi$.

We note that the curvature~\eqref{eq:sds-psi} is $\lambda^3$ times the static Coulomb field of a massless spin-two field in flat space, as required by the conformal invariance of the spin-two equation. It is of type D. It is not aligned with any constant null vector $S^i$, so by~\eqref{P-flat-Box} it cannot be written as $\mathcal P\phi$ with a rank-one potential $\phi^{ij}=\lambda^3S^iS^j f$.

\paragraph{The first-order constraint.} As emphasised in Section~\ref{sec:sourced-equations}, a solution of the second-order equation must also satisfy the first-order Bianchi equation~\eqref{eq:sourced-first-order-bianchi}, $d_2^*\psi=d_1j$. The second-order equation alone is weaker. For example, the rescaled field $\hat\psi^{ij}=c^ix^j+c^jx^i-\frac23\delta^{ij}c^kx^k$, with constant $c^i$, is harmonic and static, so it can be added to~\eqref{eq:sds-psi} without changing~\eqref{eq:sds-curvature-wave}. However, it has $d_2^*\psi\neq0$, so it is excluded by the first-order constraint.

For the curvature~\eqref{eq:sds-psi} the constraint holds, including at the worldline. To see this, it is convenient to compute both sides for $\psi^{ij}=m\,t^3D_{ij}\Phi$ with an arbitrary function $\Phi(x)$, and for the source~\eqref{eq:sds-j} with $\delta^{(3)}(x)$ replaced by $-\frac1{4\pi}\nabla^2\Phi$. One finds
\begin{equation}\label{eq:sds-constraint}
\big(d_2^*\psi\big)^i=\big(d_1j\big)^i=\frac{m\,t^3}{3}\Big(2\im\,\partial_i\nabla^2\Phi\;dt+\epsilon^{ijk}\,\partial_k\nabla^2\Phi\;dx^j\Big),
\end{equation}
so the constraint holds identically in $\Phi$. Since this is an identity between differential operators acting on $\Phi$, it continues to hold when $\Phi$ is a distribution. Taking $\Phi=\rho^{-1}$, so that $\nabla^2\Phi=-4\pi\delta^{(3)}(x)$, gives the statement for the point source.

Formula~\eqref{eq:sds-constraint} has a simple interpretation. For a static curvature $\psi=t^3\hat\psi$, the $dt$ component of $d_2^*\psi$ is proportional to the divergence $\partial_j\hat\psi^{ij}$, and the $dx^j$ components to the curl $\epsilon^{ikl}\partial_k\hat\psi^{lj}$. For $\hat\psi=m\,D_{ij}\Phi$ one has $\partial_jD_{ij}\Phi=\frac23\partial_i\nabla^2\Phi$, and the curl is produced by the trace subtraction in $D_{ij}$ alone. Away from the worldline, both vanish for $\Phi=\rho^{-1}$: the Coulomb field is divergence free and curl free, which is the vacuum Bianchi identity. At the worldline, the constraint takes the form of a Gauss law,
\[
\partial_j\hat\psi^{ij}=-\frac{8\pi m}{3}\,\partial_i\delta^{(3)}(x),
\]
the spin-two analogue of $\nabla\cdot E=4\pi q\,\delta^{(3)}$. As in electrostatics, the strength of the Coulomb field is thus fixed by the mass of the source through a first-order equation. The retarded solution therefore satisfies the constraint, consistent with the expected propagation of the constraint discussed in Section~\ref{sec:sourced-equations}.

\subsection{The Potential}\label{subsec:sds-potential}

We now follow the metric reconstruction procedure (\ref{eq:sourced-reconstruction-sequence}). By~\eqref{eq:sourced-flat-potential-reconstruction}, with $\phi=\lambda^3\chi$, the potential equation $\mathcal P\phi=\psi$ becomes 
\[
\hat\Box\chi^{ij}=-2t^{-1}\psi^{ij}=-2m\,t^2Y^{ij}/\rho^3. 
\]
The ansatz $\chi=t^2A(x)+B(x)$ reduces it to $\nabla^2A=-2mY/\rho^3$ and $\nabla^2B=2A$. Using 
\[
\nabla^2(\rho^kY^{ij})=\big(k(k+1)-6\big)\rho^{k-2}Y^{ij},
\]
one finds
\begin{equation}\label{eq:sds-potential}
\phi^{ij}=\frac m3\,t^3\,Y^{ij}\Big(\frac{t^2}{\rho}-\frac\rho2\Big).
\end{equation}
Equivalently, using $D_{ij}\rho=-\frac13Y^{ij}/\rho$ and $D_{ij}\rho^3=\rho\,Y^{ij}$,
\begin{equation}\label{eq:sds-potential-D}
\chi^{ij}=t^{-3}\phi^{ij}=-m\,D_{ij}\Big(t^2\rho+\frac{\rho^3}{6}\Big).
\end{equation}
Since $\nabla^2\rho=2/\rho$ and $\nabla^2\rho^3=12\rho$ hold in the sense of distributions, and $D_{ij}$ commutes with $\nabla^2$, this solves the potential equation distributionally, including at the origin.

We note that all five components of the potential are used. In particular, the potential is not of rank one. The $S^i S^j$-aligned potentials that are expected to suffice for the radiative perturbations of Section~\ref{sec:example-ds} cannot reproduce the curvature of a point mass, while the general five-component potential does so with a field that is regular away from the worldline.

\subsection{The Reconstructed Metric}\label{subsec:sds-metric}

Away from the worldline the term $j$ vanishes, and $\sigma=d_1^*d_3\phi$. We compute the metric in three steps, again with $\Lambda=3$.

\paragraph{The connection.} By~\eqref{eq:sourced-flat-potential-reconstruction}, $a^i_\mu=t^{-1}\sd^j_\mu{}^\nu\partial_\nu\chi^{ij}$, with $\chi^{ij}$ given by \eqref{eq:sds-potential-D}. A short computation gives
\begin{equation}\label{eq:sds-rec-connection}
a^i=\frac{2\im m}{3\rho}\Big(\frac{2(t^2-\rho^2)}{t\rho}\,n^i\,dt+Y^{ij}dx^j\Big)+\frac{2m}{3}\,\frac{t^2-\rho^2}{t\rho^2}\,\epsilon^{ijk}n^k\,dx^j .
\end{equation}

\paragraph{The curvature of the connection.} Decompose $d^Aa^i=X^{ij}\sd^j+\theta^{ij}\bar\sd^j$, with $X^{ij}=\frac14\sd^{j\mu\nu}(d^Aa^i)_{\mu\nu}$ and $\theta^{ij}=\frac14\bar\sd^{j\mu\nu}(d^Aa^i)_{\mu\nu}$. The symmetric tracefree part of $X$ is $\psi$ of~\eqref{eq:sds-psi}, as guaranteed by $d_2d_3\phi=\mathcal P\phi=\psi$. The trace of $X$ vanishes, and its antisymmetric part,
\[
X^{[ij]}=\frac{4\im m}{3}\,\frac{t^2-\rho^2}{\rho^2}\,\epsilon^{ijk}n^k ,
\]
is an internal rotation, which does not contribute to the metric. The anti-self-dual part is
\begin{equation}\label{eq:sds-theta}
\theta^{ij}=p\,\delta^{ij}+q\,n^in^j+\im r\,\epsilon^{ijk}n^k,\qquad
p=-\frac{mt(4\rho^2-t^2)}{3\rho^3},\quad q=\frac{mt(4\rho^2-3t^2)}{3\rho^3},\quad r=\frac{2m(t^2+\rho^2)}{3\rho^2}.
\end{equation}
Thus $\sigma=d_1^*a=(\mathbb I-P_5)d^Aa$ has, in the decomposition~\eqref{eq:reality-decomposition}, $s=0$ and the matrix $\theta$ above. Since $p$, $q$, $r$ are real, $\theta$ is Hermitian, and by~\eqref{eq:linearised-metric-reality} the metric is real.

\paragraph{The metric.} Here we compute the metric perturbation $\delta g_{\mu\nu}$ as produced by $\sigma$ computed in the previous step. By Appendix~\ref{apx:metric-reality}, $P_{10}(\sigma)=s\,g+\theta^{ij}K^{ij}$ with $K^{ij}_{\mu\nu}=\sd^i_{(\mu}{}^\rho\bar\sd^j_{\nu)\rho}$. In our coordinates, with symmetrised products,
\[
t^2K^{ij}=\delta^{ij}\big(dt^2+dx^kdx^k\big)-2\,dx^idx^j-2\im\,\epsilon^{ijk}\,dt\,dx^k .
\]
Contracting with~\eqref{eq:sds-theta}, and using $\operatorname{tr}\theta=3p+q=-8mt/(3\rho)$ and $\epsilon^{ijk}\theta^{ij}=2\im r\,n^k$,
\[
\delta g=\frac1{t^2}\Big[-\frac{8mt}{3\rho}\big(dt^2+dx^kdx^k\big)-2p\,dx^kdx^k-2q\,(n^kdx^k)^2+4r\,n^k\,dt\,dx^k\Big].
\]
Collecting terms gives
\begin{equation}\label{eq:sds-reconstructed-metric}
\delta g_{tt}=-\frac{8m}{3t\rho},\qquad
\delta g_{ti}=\frac{4m}{3}\,\frac{(t^2+\rho^2)\,n_i}{t^2\rho^2},\qquad
\delta g_{ij}=-\frac{2m}{3t}\Big[\frac{t^2}{\rho^3}\big(\delta_{ij}-3n_in_j\big)+\frac4\rho\,n_in_j\Big].
\end{equation}
This metric perturbation is tracefree, which is as expected, since $d_3\phi$ has no four-dimensional component (Section~\ref{subsec:gauge}). It is also real without any correction: no homogeneous potential $\phi_0$ has to be added to satisfy the reality condition of Section~\ref{subsec:version-II}. We now compare it with the known solution.

\subsection{The McVittie Form of the Schwarzschild--de Sitter Metric}\label{subsec:mcvittie}

The Schwarzschild--de Sitter solution can be written in coordinates adapted to the flat slicing of de Sitter space, in which it takes the form found by McVittie~\cite{McVittie1933}:
\begin{equation}\label{eq:mcvittie}
ds^2=-\left(\frac{1-\frac{m}{2a(T)r}}{1+\frac{m}{2a(T)r}}\right)^{\!2}dT^2+a(T)^2\Big(1+\frac{m}{2a(T)r}\Big)^{4}\big(dr^2+r^2d\Omega^2\big),\qquad a(T)=e^{HT}.
\end{equation}
Here $r$ is the comoving isotropic radius. For a general scale factor $a(T)$ the metric~\eqref{eq:mcvittie} describes a mass embedded in a spatially flat Friedmann--Lema\^itre--Robertson--Walker universe. For the exponential scale factor it is an exact vacuum solution with $\Lambda=3H^2$, and it is the Schwarzschild--de Sitter metric in these coordinates; see~\cite{Kaloper:2010mcvittie} for a discussion. For $m=0$ it reduces to de Sitter space in the flat slicing, and the worldline $r=0$ is a comoving geodesic.

We set $H=1$, so that $\Lambda=3$, and introduce the conformal time $t=e^{-T}>0$, so that $a=1/t$ and the background is the metric $g=t^{-2}(-dt^2+dx^idx^i)$ used above, with $\rho=r=|x|$. To first order in $m$ one has $-g_{TT}=1-2m/(a\rho)$ and $g_{rr}=a^2(1+2m/(a\rho))$, so the metric perturbation is $\delta g=\frac{2m}{a\rho}\,(dT^2+a^2dx^idx^i)$. In terms of $t$ this is
\begin{equation}\label{eq:sds-linearised}
\delta g_{\rm SdS}=\frac{2m}{t\rho}\,\big(dt^2+dx^idx^i\big),\qquad \rho=|x|.
\end{equation}

\subsection{Comparison with Linearised Schwarzschild--de Sitter}\label{subsec:sds-comparison}

We now compute the chiral description of the metric perturbation~\eqref{eq:sds-linearised} step by step, and compare it with the objects arising from the reconstruction.

\paragraph{The 2-form perturbation.} The 2-form perturbation encoding~\eqref{eq:sds-linearised}, in the representative without internal rotation, is $\sigma=U(\delta g_{\rm SdS})$, with $U$ as in~\eqref{eq:source-U-map}. Since $dt^2+dx^idx^i=t^2g+2\,dt\otimes dt$ and $U(g)=\sd$, only $U(dt\otimes dt)$ needs to be computed. For $q=dt\otimes dt$ one has $q_\mu{}^\rho=-t^2\delta_\mu^t\delta^\rho_t$, and only the components $\sd^i_{tj}=\im t^{-2}\delta^i_j$ of the triple contribute, giving $U(dt\otimes dt)^i=-\frac\im2\,dt\wedge dx^i$. Hence
\[
\sigma^i=\frac{2m}{t\rho}\Big(t^2\sd^i-\im\,dt\wedge dx^i\Big),
\]
in which the terms $dt\wedge dx^i$ cancel, leaving
\begin{equation}\label{eq:sds-sigma}
\sigma^i=-\frac{m}{t\rho}\,\epsilon^{ijk}dx^j\wedge dx^k=\frac{mt}{\rho}\big(\sd^i+\bar\sd^i\big).
\end{equation}
In the decomposition~\eqref{eq:reality-decomposition} this is $s=mt/\rho$ and $\theta^{ij}=(mt/\rho)\,\delta^{ij}$, with no internal rotation. Both are real, and $\theta$ is Hermitian, as required for a real metric by~\eqref{eq:linearised-metric-reality}. The 2-form perturbation is purely spatial and isotropic.

\paragraph{The connection.} The compatible connection is $a=d_1\sigma=\im J_1^{-1}\star d^A\sigma$, which we compute in three steps. First, $A^j\wedge\sigma^k=\frac{2\im m}{t^2\rho}\,\delta^{jk}\,d^3x$, with $d^3x=dx^1\wedge dx^2\wedge dx^3$, so that $\epsilon^{ijk}A^j\wedge\sigma^k=0$ and $d^A\sigma=d\sigma$:
\[
d^A\sigma^i=\frac{m}{t^2\rho}\,dt\wedge\epsilon^{ijk}dx^j\wedge dx^k+\frac{2m\,x^i}{t\rho^3}\,d^3x .
\]
Second, taking the Hodge dual,
\[
\star d^A\sigma^i=-\frac{2m}{\rho}\Big(\frac{t\,x^i}{\rho^2}\,dt+dx^i\Big).
\]
Third, with $J_1(\theta)^i=\epsilon^{ijk}\sd^j\cdot\theta^k$ one finds
\[
J_1(\star d^A\sigma)^i=-\frac{4m}{\rho}\,dx^i-\frac{2\im m\,t}{\rho^3}\,\epsilon^{ijk}x^k\,dx^j ,
\]
and $J_1^{-1}=\frac12(J_1-\mathbb I)$ (Appendix~\ref{apx:torsion-free-sol}) gives $a=\frac\im2(J_1-\mathbb I)\star d^A\sigma$. Using $x^i/\rho^3=-\partial_i\rho^{-1}$,
\begin{equation}\label{eq:sds-connection}
a^i=-\im m\Big(t\,\partial_i\frac1\rho\,dt+\frac1\rho\,dx^i\Big)-m\,t\,\epsilon^{ijk}\,\partial_k\frac1\rho\,dx^j .
\end{equation}
This differs from the reconstructed connection~\eqref{eq:sds-rec-connection}, as expected, since the two metrics turn out to differ by a gauge transformation, see~\eqref{eq:sds-gauge} below.

\paragraph{The curvature.} We compute $d^Aa^i=da^i+\epsilon^{ijk}A^j\wedge a^k$ term by term. Writing $\partial_i\partial_j\rho^{-1}=\partial_{ij}\rho^{-1}$,
\begin{align*}
da^i&=\im m\,t\,\partial_{ij}\rho^{-1}\,dt\wedge dx^j-\im m\,\partial_j\rho^{-1}\,dx^j\wedge dx^i-m\,\epsilon^{ijk}\partial_k\rho^{-1}\,dt\wedge dx^j-m\,t\,\epsilon^{ijk}\partial_{kl}\rho^{-1}\,dx^l\wedge dx^j,\\
\epsilon^{ijk}A^j\wedge a^k&=m\,\epsilon^{ijk}\partial_k\rho^{-1}\,dt\wedge dx^j+\im m\,\partial_j\rho^{-1}\,dx^j\wedge dx^i-\frac{m}{t\rho}\,\epsilon^{ijk}dx^j\wedge dx^k .
\end{align*}
The terms with first derivatives of $\rho^{-1}$ cancel, and the last term is $\sigma^i$. Away from the worldline $\partial_{kl}\rho^{-1}$ is symmetric and tracefree, and for such matrices $\epsilon^{ijk}S_{kl}\,dx^j\wedge dx^l=-\frac12S_{ij}\,\epsilon^{jkl}dx^k\wedge dx^l$. The remaining two terms therefore combine into $m\,t\,\partial_{ij}\rho^{-1}\big(\im\,dt\wedge dx^j-\frac12\epsilon^{jkl}dx^k\wedge dx^l\big)=m\,t^3\,\partial_{ij}\rho^{-1}\,\sd^j$. We thus find
\begin{equation}
d^Aa^i=\psi^{ij}\sd^j+\frac\Lambda3\,\sigma^i,
\end{equation}
with $\Lambda=3$, and with the linearised self-dual Weyl curvature given by
\begin{equation}\label{eq:sds-psi-SdS}
\psi^{ij}=d_2(a)^{ij}=m\,t^3\,\partial_i\partial_j\frac1\rho=m\,t^3\,\frac{Y^{ij}}{\rho^3}.
\end{equation}
This is the curvature~\eqref{eq:sds-psi} obtained from the source in Section~\ref{subsec:sds-source}: the linearised Schwarzschild--de Sitter metric and the reconstructed metric have the same self-dual curvature.

Since both metrics solve the linearised Einstein equation with the same source and have the same self-dual curvature, and the reconstructed metric is real, local uniqueness implies that they differ by a gauge transformation. Explicitly, the reconstructed metric~\eqref{eq:sds-reconstructed-metric} and~\eqref{eq:sds-linearised} are related by a diffeomorphism,
\begin{equation}\label{eq:sds-gauge}
\delta g=\delta g_{\rm SdS}+\mathcal L_\xi g,\qquad
\xi=\frac m3\Big[\frac{7t^2}{\rho}\,\partial_t+t\Big(\frac4\rho-\frac{t^2}{\rho^3}\Big)x^i\partial_i\Big],
\end{equation}
which we have verified by direct computation. 

\section{Discussion}\label{sec:discussion}

The purpose of this section is to compare the first-order reconstruction developed in this paper with the standard Teukolsky--Wald reconstruction, and to identify as precisely as possible what is simplified in the new scheme, and at what cost. Any reconstruction scheme has to address three issues: (a) how a metric perturbation with a prescribed curvature is obtained from a potential; (b) how sources are accommodated; and (c) how the reality of the reconstructed metric perturbation is achieved. We first review how the standard construction on Kerr deals with each of these, in Section~\ref{subsec:reality-comparison}. We then describe two versions of the first-order reconstruction. Version~I, in Section~\ref{subsec:version-I}, is a direct translation of the standard construction, including its treatment of sources. Version~II, in Section~\ref{subsec:version-II}, is the reconstruction of Section~\ref{subsec:sourced-potential-reconstruction}; it is obtained from version~I by a particular choice of the source of the potential equation, for which the treatment of sources simplifies. Section~\ref{subsec:simplification} summarises what is gained and what is not. The example of Section~\ref{sec:example-static-mass}, a static mass in de Sitter space, illustrates version~II explicitly, and we refer to it below where it bears on the general discussion. The remaining subsections discuss further aspects of the construction, as well as possible extension to Kerr.

\subsection{Metric Reconstruction: the Teukolsky--Wald Framework}\label{subsec:reality-comparison}

We review the standard construction on Kerr, organised around the three issues listed above. The material in this subsection is standard; see~\cite{Chrzanowski1975,KegelesCohen1979,Wald:1978vm,Green:2020sourced,Black_hole_pert_Pound_2021} and references therein.

\paragraph{Helicities and curvature.} In the complexified linear theory a metric perturbation has two independent linearised Weyl curvatures. The self-dual part is represented in this paper by the symmetric tracefree matrix $\psi^{ij}$, equivalently by the perturbation of the Weyl spinor $\psi_{ABCD}$. The anti-self-dual part, which does not appear explicitly in our construction, we denote by $\tilde\psi$; in spinor language it is the primed spinor $\tilde\psi_{A'B'C'D'}$. For a real Lorentzian perturbation these two curvatures are not independent: $\tilde\psi=\bar\psi$, with the conjugation conventions fixed by the background. A complex perturbation with $\psi=0$ and $\tilde\psi\neq0$, or vice versa, describes a single graviton helicity. Such a perturbation cannot be real, since reality would force both curvatures to vanish.

In Newman--Penrose language one uses a null tetrad $(l,n,m,\bar m)$, with $l$ and $n$ real and $m$ complex. The five Weyl scalars are the components of the linearised Weyl tensor along this tetrad\footnote{In the Kerr literature these are denoted $\Psi_0,\dots,\Psi_4$, and the Hertz potential $\Psi_H$. We use lower case to distinguish the perturbation from the background curvature $\Psi$.}, for example
\[
\psi_0=C_{abcd}\,l^am^bl^cm^d,\qquad \psi_4=C_{abcd}\,n^a\bar m^bn^c\bar m^d,
\]
and they are the five components of the self-dual Weyl spinor. The corresponding anti-self-dual components are obtained by exchanging $m$ and $\bar m$; for instance $\tilde\psi_0=C_{abcd}\,l^a\bar m^bl^c\bar m^d$. Because $m$ is complex, the Weyl scalars are complex numbers even for a real metric perturbation. For a real perturbation, $\tilde\psi_0=\bar\psi_0$, and similarly for the other components. A single complex function such as $\psi_0$ therefore encodes the two real degrees of freedom of a real gravitational perturbation: roughly speaking, its real and imaginary parts carry the two polarisations. We will use repeatedly the fact that $\psi_0$ is linear over the complex numbers in the metric perturbation: $\psi_0[h_1+ch_2]=\psi_0[h_1]+c\,\psi_0[h_2]$ for complex $c$. By contrast, $\psi_0[\bar h]$ is not the complex conjugate of $\psi_0[h]$; rather, $\psi_0[\bar h]=\overline{\tilde\psi_0[h]}$, since complex conjugation of the metric perturbation exchanges the roles of $m$ and $\bar m$.

\paragraph{The standard construction on Kerr.} In the Teukolsky--Wald framework, the reconstruction of a real metric perturbation proceeds in the following steps~\cite{Chrzanowski1975,KegelesCohen1979,Wald:1978vm,Green:2020sourced,Berens:2024kerr}. Normalisations differ between references; numerical factors below refer to the conventions of~\cite{Berens:2024kerr}. We first describe the construction for source-free perturbations, or in a region where the source vanishes. The modifications required in the presence of sources are described in the paragraph on sources and the corrector below.
\begin{enumerate}
\item One solves the source-free Teukolsky equation $\mathcal O_0\psi_0=0$ for one of the extreme Weyl scalars, say $\psi_0$. This is a single complex function, and it is the Weyl scalar of the real, physical metric perturbation that one wants to construct. Causality enters at this step: for a physical problem one takes the retarded solution. Here and below we write $\mathcal T_0$ for the map $h\mapsto\psi_0[h]$, and $\mathcal O_0$, $\mathcal S_0$ for the Teukolsky operator of spin weight $+2$ and the corresponding decoupling map, so that Wald's identity reads $\mathcal S_0\mathcal G=\mathcal O_0\mathcal T_0$; there is an analogous triple $\mathcal T_4,\mathcal O_4,\mathcal S_4$ for $\psi_4$. With a source $J$ the equation becomes $\mathcal O_0\psi_0=\mathcal S_0J$, and steps~2 and~3 below then have to be modified.
\item One constructs metric perturbations from Hertz potentials. Let $\psi_H$ be a complex solution of the adjoint Teukolsky equation $\mathcal O_0^* \psi_H=0$. In the ingoing radiation gauge, the adjoint Teukolsky equation is the Teukolsky equation of spin weight $-2$, that is, the equation satisfied by $\zeta^4\psi_4$, with $\zeta$ referred to as the Killing spinor coefficient in~\cite{Black_hole_pert_Pound_2021}. The Hertz potential is, however, not the $\psi_4$ of the reconstructed metric: it merely satisfies the same equation. Both extreme Weyl scalars $\psi_0$ and $\psi_4$ of the reconstructed metric are fourth-order expressions in $\psi_H$; for $\psi_0$ see~\eqref{psi0-psiH} below. Wald's identity guarantees that
\[
h^c:=\mathcal S_0^*\psi_H
\]
solves the source-free linearised Einstein equation. This solution is complex. Since the linearised Einstein operator is real, its complex conjugate $\overline{h^c}$ is also a solution, and so is the real metric perturbation
\[
h=h^c+\overline{h^c}=2\,\mathrm{Re}\,h^c .
\]
It is this real metric $h$ that is used as the reconstructed metric. Reality is thus imposed by construction, at this step.
\item One chooses the Hertz potential so that the real metric $h$ of step~2 has the prescribed $\psi_0$ of step~1. To understand this, we need to analyse the role of the two parts of $h=h^c+\overline{h^c}$. The operator $\mathcal T_0$ is linear over the complex numbers. For the complex metric $h^c=\mathcal S_0^*\psi_H$ one has
\[
\mathcal T_0 h^c=\psi_0[h^c]=0 .
\]
This is a local identity: it is a property of the reconstruction operator $\mathcal S_0^*$ in the ingoing radiation gauge. It holds at every point where $h^c$ is given by the Hertz formula, whether or not sources are present elsewhere, and it involves neither causality nor boundary conditions. The reason is that in the ingoing radiation gauge the Hertz potential is aligned with the principal spinor $o_A$ of the background. This alignment forces $\psi_0,\dots,\psi_3$ of $h^c$ to vanish for any function $\psi_H$, while $\psi_4[h^c]$ vanishes once $\psi_H$ satisfies its field equation;\footnote{On Kerr, $\psi_4[h^c]$ does not vanish on shell. What remains is a term proportional to the black-hole mass times the time derivative of $\psi_H$, see~\cite{Black_hole_pert_Pound_2021}.} the flat-space argument is given in Appendix~\ref{apx:hertz-alignment}. 

While $\psi_0[h^c]=0$ vanishes identically as discussed, this is no longer the case for $\psi_0[\overline{h^c}]$. Indeed, for the complex conjugate metric, 
\begin{equation}\label{psi0-psiH}
\mathcal T_0\,\overline{h^c}=\psi_0\big[\overline{h^c}\big]=\tfrac14\,l^4\,\bar\psi_H\neq0 .
\end{equation}
where $l$ now denotes the derivative along the principal null direction~\cite{Berens:2024kerr}, and $\bar\psi_H$ is the complex conjugate of the Hertz potential $\psi_H$. Given the complex function $\psi_0$, this is a fourth-order equation for the complex function $\psi_H$; the counting matches, one complex function on each side. Inverting this relation, i.e.\ solving a fourth-order ordinary differential equation along the principal null rays, determines $\psi_H$ up to elements of the kernel of $l^4$, which are fixed by boundary or causality conditions~\cite{Ori2003}.
\end{enumerate}

\paragraph{Two equations for the Hertz potential.} The Hertz potential has to satisfy two equations: the adjoint Teukolsky equation $\mathcal O_0^*\psi_H=0$ of step~2, which guarantees that $h$ is a solution, and the inversion relation~\eqref{psi0-psiH} of step~3, which fixes its curvature. A solution of~\eqref{psi0-psiH} alone is determined only up to the kernel of $l^4$, that is, up to four complex functions that are constant along the principal null rays, and it does not in general solve the adjoint Teukolsky equation. These integration functions must therefore be chosen so that it does. That this is always possible for a source-free $\psi_0$ is a result of the literature, not an additional assumption. In the frequency domain, the inversion relation is solved algebraically, mode by mode, by imposing the adjoint Teukolsky equation at the same time. The Teukolsky--Starobinsky identities then guarantee that the resulting $\psi_H$ satisfies both equations~\cite{Ori2003,Berens:2024kerr}. More generally, the existence of a Hertz potential with the required properties follows from the decomposition theorem of~\cite{Green:2020sourced} recalled below, in which the corrector and the source of the equation for $\psi_H$ are constructed from the stress-energy tensor and can be taken to vanish when it does. The same compatibility question arises in version~I of the first-order framework, see Section~\ref{subsec:version-I}.

\paragraph{The Teukolsky--Starobinsky identities.} The Teukolsky--Starobinsky identities~\cite{TeukolskyPress1974,TheMathematicaChandr1998} are statements about the two Teukolsky equations: certain fourth-order operators map solutions of the spin weight $-2$ equation to solutions of the spin weight $+2$ equation, and vice versa. They can therefore be applied to two different pairs of fields. The first is $\psi_0$ and $\zeta^4\psi_4$ of a single solution of the linearised Einstein equations, which is the usual form of the identities. The second, which is the one relevant for reconstruction, is $\psi_0$ and the Hertz potential $\psi_H$, which satisfies the spin weight $-2$ equation without being the $\psi_4$ of the reconstructed metric. 

The nonvanishing in (\ref{psi0-psiH}) is why the Teukolsky--Starobinsky identities are needed. Indeed, $\overline{h^c}$ is itself a solution of the source-free linearised Einstein equation, $\mathcal G\,\overline{h^c}=0$, because $\mathcal G$ is real. Applying Wald's identity $\mathcal S_0\mathcal G=\mathcal O_0\mathcal T_0$ to $\overline{h^c}$ gives $\mathcal O_0\,\mathcal T_0\,\overline{h^c}=0$, that is,
\[
\mathcal O_0\big(l^4\,\bar\psi_H\big)=0\qquad\text{whenever}\qquad\mathcal O_0^*\psi_H=0 .
\]
In words: the fourth-order operator $l^4$, combined with complex conjugation, maps solutions of the adjoint Teukolsky equation, i.e.\ of the spin weight $-2$ equation, to solutions of the spin weight $+2$ Teukolsky equation satisfied by $\psi_0$.  

\paragraph{Sources and the corrector.} In steps~2 and~3 above, the Hertz potential solves the source-free adjoint Teukolsky equation, and the reconstructed metric solves the source-free linearised Einstein equation. Suppose now that the metric perturbation is to solve $\mathcal G h=J$ for a given conserved stress-energy tensor $J$. The natural generalisation is to let the Hertz potential solve a sourced equation, $\mathcal O_0^*\psi_H=\eta$. The adjoint identity~\eqref{teuk-adj-intr} then gives
\[
\mathcal G\,\mathcal S_0^*\psi_H=\mathcal T_0^*\mathcal O_0^*\psi_H=\mathcal T_0^*\eta ,
\]
so that the reconstructed metric solves the Einstein equation with source $\mathcal T_0^*\eta$, together with its complex conjugate after the real part is taken. The difficulty is that a general stress-energy tensor $J$ is not of this form: the image of $\mathcal T_0^*$ is restricted, and no choice of $\eta$ reproduces a generic source. Equivalently, a metric perturbation in the radiation gauge cannot in general solve the Einstein equation with a generic source; for point particles this is the origin of the well-known string-like singularities of radiation-gauge metric perturbations, see~\cite{Black_hole_pert_Pound_2021}.

This difficulty was resolved by Green, Hollands and Zimmerman~\cite{Green:2020sourced}. They showed that a solution of $\mathcal G h=J$ can be decomposed, with their normalisations, as
\[
h=\mathcal L_\xi g+\dot g+x+\mathrm{Re}\,\mathcal S_0^*\psi_H ,
\]
the sum of a pure-gauge part, a zero mode $\dot g$ (a perturbation of the mass and angular momentum), a \emph{corrector tensor} $x$, and a part reconstructed from a Hertz potential, which solves the spin weight $-2$ Teukolsky equation with a source. The corrector absorbs the part of the source that the Hertz-potential part cannot produce. Its tetrad components, and the source $\eta$ of the equation for the Hertz potential, are obtained by integrating decoupled ordinary differential equations involving the stress-energy tensor along the outgoing principal null geodesics. These integrations are much simpler than solving partial differential equations, but they are nonlocal: for a point particle, the corrector is supported on semi-infinite strings extending from the worldline to infinity~\cite{Green:2020sourced}. Wald's uniqueness theorem~\cite{Wald1973}, which states that a perturbation with vanishing $\psi_0$ is of the zero-mode-plus-gauge form, then guarantees that the decomposition captures the physical metric perturbation.

\paragraph{Summary of the standard construction.} In the standard construction, (a) a prescribed curvature $\psi_0$ is reached by inverting the fourth-order relation~\eqref{psi0-psiH}; (b) sources are accommodated by a sourced equation for the Hertz potential together with the corrector tensor, both obtained by integrations along null geodesics, the corrector being nonlocal; and (c) reality is imposed by construction, by taking the real part.

\subsection{Metric Reconstruction in the First-Order Framework: Version I}\label{subsec:version-I}

We now turn to the first-order framework, and describe version~I, the direct analogue of the standard construction. 

We start with a general remark. On the constant-curvature backgrounds of this paper the background Weyl curvature vanishes, and the two helicities decouple exactly. The linearised Weyl curvature is then gauge invariant. Moreover, for a given source, two solutions of the linearised Einstein equations with the same pair of curvatures $(\psi,\tilde\psi)$ differ by a source-free perturbation that preserves constant curvature with the given $\Lambda$, and such a perturbation is locally pure gauge. Thus, locally and modulo gauge, a complex solution is determined by the pair $(\psi,\tilde\psi)$. Each of the two curvatures satisfies its own spin-two equation, with the same source, and nothing couples them. 

It will be useful to state the equations for $\tilde{\psi}$. Let $\bar d_1$, $\bar d_2$ denote the operators obtained from $d_1$, $d_2$ by replacing $\sd^i$ and $A^i$ by their anti-self-dual counterparts. The linearised Bianchi identity holds for every perturbation, real or complex. For a perturbation of an Einstein background with vanishing Weyl curvature it splits into the self-dual equation $d_2^*\psi=d_1j$ of~\eqref{eq:sourced-first-order-bianchi} and its counterpart
\begin{equation}\label{asd-bianchi}
\bar d_2^*\tilde\psi=\bar d_1\bar j ,
\end{equation}
where $\bar j$, the complex conjugate of $j$, encodes the same real stress-energy tensor through the anti-self-dual triple. In spinor language this is $\nabla^{A}{}_{A'}\tilde\psi^{A'}{}_{B'C'D'}$ equal to the derivative of the stress-energy tensor. We call it the anti-self-dual spin-two equation.

\paragraph{Version I: Wald-type reconstruction.} In the first version, the potential solves the homogeneous equation $\mathcal P\phi=0$, as in Section~\ref{sec:conformal-reconstruction}. The complex perturbation $d_1^*d_3\phi$ then plays the role of the complex Hertz metric $h^c=\mathcal S_0^*\psi_H$ of the standard construction. Its properties are as follows.

First, the self-dual curvature vanishes. The curvature-extraction operator of our framework is $\mathcal T=d_2d_1$, and by the principal identity~\eqref{identity-intr}
\[
\mathcal T\,d_1^*d_3\phi=d_2\big(\mathbb I+d_2^*d_2\big)d_3\phi=\big(\mathbb I+d_2d_2^*\big)\mathcal P\phi .
\]
For a solution of $\mathcal P\phi=0$ the right-hand side vanishes, so the complex perturbation $d_1^*d_3\phi$ has $\psi=0$. This uses the potential equation, and it holds for a general potential $\phi^{ij}$. Note that $\mathcal T$ extracts all five components of the self-dual Weyl curvature at once, and is thus the first-order counterpart of the collection $(\mathcal T_0,\dots,\mathcal T_4)$ of the usual approach. On constant-curvature backgrounds all five components obey decoupled equations.

Second, the anti-self-dual curvature of $d_1^*d_3\phi$ is in general nonzero. We denote it by $\tilde\psi=\mathcal K\phi$, where $\mathcal K$ is the linear map from the potential to the anti-self-dual curvature of $d_1^*d_3\phi$. Since this metric perturbation involves two derivatives of $\phi$, and the curvature two further derivatives, $\mathcal K$ is a fourth-order operator. Since $d_1^*d_3\phi$ solves the source-free linearised Einstein equation, the anti-self-dual Bianchi equation (\ref{asd-bianchi}) gives $\bar d_2^*\mathcal K\phi=0$. Thus $\mathcal K$ maps solutions of $\mathcal P\phi=0$ to solutions of the homogeneous anti-self-dual spin-two equation, just as $l^4$ maps solutions of the adjoint Teukolsky equation to solutions of the Teukolsky equation in the standard construction.

Third, the perturbation $d_1^*d_3\phi$ is therefore of a single helicity, and hence necessarily complex. The explicit de Sitter metric~\eqref{eq:ds-metric}, obtained from the rank-one potentials $\phi^{ij}=\lambda^3S^iS^jf$ with $S^i$ null, shows this concretely: it takes values in a totally null complex two-plane.

Fourth, real solutions are obtained by taking the real part of the metric. Since $\sd^i$ is complex, a real metric is encoded by a complex 2-form, and the componentwise real part $\frac12(\sigma+\bar\sigma)$ is not the correct operation: complex conjugation exchanges self-dual and anti-self-dual 2-forms, so $\frac12(\sigma+\bar\sigma)$ in general violates linearised metricity~\eqref{eq:lin-metricity-cond}. The correct operation is the algebraic real-metric projection $\mathcal R$ of Appendix~\ref{apx:metric-reality}, defined in~\eqref{eq:two-form-real-projection}. It satisfies
\[
P_{10}(\mathcal R\sigma)=\mathrm{Re}\,P_{10}(\sigma),
\]
see~\eqref{eq:real-projection-metric}. Equivalently, $\mathcal R\sigma=U(\mathrm{Re}\,P_{10}\sigma)$, with $U$ as in~\eqref{eq:source-U-map}: $\mathcal R\sigma$ is the 2-form encoding of the real part of the metric of $\sigma$, in the representative without internal rotation.

We apply this to the complex solution $\sigma_c=d_1^*d_3\phi$, with $\mathcal P\phi=0$. The connection $d_3\phi$ has no four-dimensional component, by~\eqref{a-8}, so the metric of $\sigma_c$ is tracefree (Section~\ref{subsec:gauge}) and $s=0$ in the decomposition~\eqref{eq:reality-decomposition}. Hence
\[
\sigma_{\mathrm R}:=\mathcal R\sigma_c,\qquad
\sigma_{\mathrm R}^{\,i}=\tfrac12\big(\theta_c+\theta_c^\dagger\big)^{ij}\bar\Sigma^j,\qquad
\theta_c^{ij}=\tfrac14\bar\Sigma^{j\mu\nu}\sigma^i_{c\,\mu\nu}.
\]
Since the linearised Einstein operator on metrics has real coefficients, $P_{10}\sigma_{\mathrm R}=\mathrm{Re}\,P_{10}\sigma_c$ is again a solution. Therefore $E_\sigma\sigma_{\mathrm R}=0$, because $E_\sigma\sigma=0$ is equivalent to the linearised Einstein equation for $P_{10}\sigma$, the internal-rotation part of $\sigma$ being pure gauge (Section~\ref{subsec:gauge}); the compatible connection is $a_{\mathrm R}=d_1\sigma_{\mathrm R}$. The curvatures of the complex and real solutions are
\[
(\psi_c,\tilde\psi_c)=(0,\mathcal K\phi),\qquad
(\psi_{\mathrm R},\tilde\psi_{\mathrm R})=\Big(\tfrac12\overline{\mathcal K\phi},\ \tfrac12\mathcal K\phi\Big),
\]
where the complex conjugate of an anti-self-dual curvature is read as a self-dual one. 

To see why the anti-self-dual curvature of $\sigma_{\mathrm R}$ is still given by the same map $\mathcal K$, note first that the curvatures of a 2-form perturbation depend only on its metric $h=P_{10}\sigma$, since internal rotations leave both $\psi$ and $\tilde\psi$ unchanged (Section~\ref{subsec:gauge}). Write $\psi[h]$ and $\tilde\psi[h]$ for the self-dual and anti-self-dual parts of the linearised Weyl tensor of $h$. These maps are linear over the complex numbers. The linearised Weyl tensor has real coefficients, and complex conjugation exchanges its self-dual and anti-self-dual parts, so that $\psi[\bar h]=\overline{\tilde\psi[h]}$ and $\tilde\psi[\bar h]=\overline{\psi[h]}$. With $h_c=P_{10}\sigma_c$ and $P_{10}\sigma_{\mathrm R}=\frac12(h_c+\bar h_c)$ this gives
\[
\tilde\psi_{\mathrm R}=\tfrac12\big(\tilde\psi[h_c]+\overline{\psi[h_c]}\big)=\tfrac12\mathcal K\phi,\qquad
\psi_{\mathrm R}=\tfrac12\big(\psi[h_c]+\overline{\tilde\psi[h_c]}\big)=\tfrac12\overline{\mathcal K\phi},
\]
where we used $\psi[h_c]=0$. The anti-self-dual curvature is thus unchanged up to the factor $\frac12$ precisely because $\sigma_c$ has vanishing self-dual curvature: the conjugate part $\frac12\bar h_c$ contributes only to the self-dual curvature. This is the same mechanism as in the standard construction, where $\psi_0[h^c]=0$ and only $\overline{h^c}$ contributes to $\psi_0$ of the real metric, see~\eqref{psi0-psiH}. Thus the projection supplies both helicities. This is the counterpart of step~2 of the standard construction. 

Finally, to obtain in this way a real vacuum solution with a prescribed self-dual curvature $\psi$, one must solve the fourth-order equation $\mathcal K\phi=2\bar\psi$ for a potential with $\mathcal P\phi=0$. As for the Hertz potential, the potential must then satisfy two equations at once. The two are consistent in the sense that $\mathcal K$ maps solutions of $\mathcal P\phi=0$ to solutions of the anti-self-dual spin-two equation, which $\bar\psi$ also solves. We have not, however, shown in general that the freedom in the solution of $\mathcal K\phi=2\bar\psi$ suffices to impose $\mathcal P\phi=0$.

Version~I thus reaches a prescribed self-dual curvature only through the fourth-order equation $\mathcal K\phi=2\bar\psi$, exactly as the standard construction reaches a prescribed $\psi_0$ through the inversion of~\eqref{psi0-psiH}.

Our construction also allows general, non-aligned potentials $\phi^{ij}$. For the source-free radiative perturbations of version~I the additional components are expected to be redundant (Section~\ref{subsec:redundancy}), so this is not an essential difference. They are, however, essential in version~II: since $\mathcal P$ maps rank-one potentials to curvatures aligned with $S^iS^j$, a non-aligned curvature, such as the Coulomb field of a point mass, requires a potential with all five components, see Section~\ref{sec:example-static-mass}.

\paragraph{Version I with sources.} Version~I has a sourced form, which follows the standard construction step by step. Let the potential solve a sourced equation, $\mathcal P\phi=\eta$. Using the intertwining relation $E_\sigma d_1^*=d_1^*E_a$ and the principal identity in the form $E_ad_3=d_2^*d_2d_3=d_2^*\mathcal P$, one finds
\[
E_\sigma\, d_1^*d_3\phi=d_1^*E_a d_3\phi=d_1^*d_2^*\,\mathcal P\phi=\mathcal T^*\eta,\qquad \mathcal T^*=d_1^*d_2^* .
\]
This is the first-order form of the adjoint identity $\mathcal G\mathcal S^*=\mathcal T^*\mathcal O^*$, with $\mathcal G=E_\sigma$, $\mathcal S^*=d_1^*d_3$ and $\mathcal O^*=\mathcal P$. The perturbation $d_1^*d_3\phi$ therefore solves the sourced equation $E_\sigma\sigma=j$ if and only if $j=\mathcal T^*\eta$. This is the same restriction as in the standard construction. We expect that a general source $j$, which contains the stress-energy tensor algebraically, is not of the form $d_1^*d_2^*\eta$, i.e.\ two derivatives of a curvature-type field. Version~I with sources therefore requires, exactly as in the standard construction, a corrector: a perturbation $x$ such that $\sigma=d_1^*d_3\phi+x$ solves $E_\sigma\sigma=j$ for a suitable choice of $\eta$. The question is how complicated this corrector has to be. Version~II answers it.

\subsection{Metric Reconstruction in the First-Order Framework: Version II}\label{subsec:version-II}

Version~II is version~I with sources, with a particular choice of the source of the potential equation: the actual self-dual curvature of the solution, $\eta=\psi$. For this choice the corrector becomes local and algebraic. Indeed, the curvature satisfies the sourced first-order Bianchi equation~\eqref{eq:sourced-first-order-bianchi}, $d_2^*\psi=d_1j$. Hence
\[
\mathcal T^*\psi=d_1^*d_2^*\psi=d_1^*d_1 j=j+E_\sigma j ,
\]
so that $d_1^*d_3\phi$, with $\mathcal P\phi=\psi$, has source $j+E_\sigma j$: the correct source plus the mismatch $E_\sigma j$. The mismatch is removed by the local term $x=-j$, since $E_\sigma(-j)=-E_\sigma j$:
\[
E_\sigma\big(d_1^*d_3\phi-j\big)=j .
\]
This is exactly the reconstruction $\sigma=d_1^*a-j$ of Section~\ref{subsec:sourced-potential-reconstruction}. In version~II the corrector is therefore $x=-j$, and its metric part, $-P_{10}\,j$, is an algebraic multiple of the stress-energy tensor, see~\eqref{eq:sourced-metric-reconstruction}. It is supported where the source is supported, in contrast to the corrector of the standard construction, which for a point particle is supported on strings extending to infinity. In the example of Section~\ref{sec:example-static-mass} the reconstructed metric is regular away from the worldline, and it is related to the linearised Schwarzschild--de Sitter metric by a diffeomorphism that is also regular there, see~\eqref{eq:sds-gauge}. The corrector is local because, for $\Lambda\neq0$, the map from connections back to 2-form perturbations is algebraic, and with the choice $\eta=\psi$ the part of the source that the potential does not produce is precisely the one that this algebraic map accounts for.

A second consequence of the choice $\eta=\psi$ is that the prescribed self-dual curvature is reproduced directly: the curvature of the connection $a=d_3\phi$ is $d_2a=\mathcal P\phi=\psi$. The fourth-order inversion $\mathcal K\phi=2\bar\psi$ of version~I is replaced by the second-order equation $\mathcal P\phi=\psi$. What is no longer automatic is reality, to which we turn after listing the steps.

In detail, version~II proceeds as follows.
\begin{enumerate}
\item One determines the self-dual curvature $\psi$ of the solution one wants to construct. With a source, $\psi$ solves the sourced curvature equation~\eqref{eq:sourced-weighted-curvature-wave} together with the first-order Bianchi constraint~\eqref{eq:sourced-first-order-bianchi}. Without a source, $\psi$ is a solution of the homogeneous spin-two equation; for a real radiative solution it is nonzero. For a static point mass in de Sitter space this step is carried out in Section~\ref{subsec:sds-source}, where the retarded solution of the second-order equation is found to satisfy the first-order constraint.
\item One solves the second-order equation $\mathcal P\phi=\psi$, see~\eqref{eq:inhomogeneous-potential-equation}, which after the conformal rescaling is a componentwise flat wave equation. Its solution is unique only up to $\phi\mapsto\phi+\phi_0$ with $\mathcal P\phi_0=0$. In general, this freedom is not pure gauge, see below. 
\item One forms $a=d_3\phi$ and $\sigma=d_1^*a-j$. The resulting complex metric perturbation $\delta g_c=P_{10}\sigma$ solves the sourced linearised Einstein equation, and its self-dual curvature is exactly $\psi$. Unlike in the standard construction, one cannot add the complex conjugate at this stage, since that would change the self-dual curvature, which has already been prescribed.
\item Reality must therefore be checked. The anti-self-dual curvature of $\delta g_c$ has the form
\[
\tilde\psi_c=\mathcal K\phi-\overline{\mathcal T j},
\]
where $\mathcal K$ is the fourth-order operator of version~I, and $-\overline{\mathcal T j}$ is the contribution of the local term $-j$. Indeed, the metric part $P_{10}j$ of $j$ is real, being algebraic in the stress-energy tensor, and the anti-self-dual curvature of a real metric perturbation is the complex conjugate of its self-dual curvature, which is $\mathcal Tj=d_2d_1j$. This is a second-order expression in the stress-energy tensor. The reality condition is
\begin{equation}\label{eq:reality-condition}
\mathcal K\phi-\overline{\mathcal T j}=\bar\psi .
\end{equation}
\end{enumerate}

\paragraph{Imposing reality.} Steps 1--3 produce a complex perturbation $\delta g_c$ whose self-dual curvature is the prescribed $\psi$. What is not under control is its anti-self-dual curvature $\tilde\psi_c$, which reality requires to equal $\bar\psi$. The discrepancy can be removed using the freedom $\phi\mapsto\phi+\phi_0$ left in step~2, as follows.
\begin{enumerate}
\item \emph{The discrepancy is a free field.} Define $r=\bar\psi-\tilde\psi_c$. Complex conjugation exchanges the two curvatures, so $\bar\psi$ is the anti-self-dual curvature of $\overline{\delta g_c}$. Since the background, the linearised Einstein operator and the source are real, $\delta g_c$ and $\overline{\delta g_c}$ solve the same sourced equation. Their anti-self-dual curvatures therefore solve the same sourced spin-two equation, and $r$ solves the homogeneous one.
\item \emph{The freedom in step 2 shifts only $\tilde\psi_c$.} Adding $\phi_0$ with $\mathcal P\phi_0=0$ adds to $\delta g_c$ the version~I perturbation built from $\phi_0$. Its self-dual curvature vanishes, so $\psi$ and the source are unchanged, while its anti-self-dual curvature is $\mathcal K\phi_0$. Thus $r\mapsto r-\mathcal K\phi_0$.
\item \emph{Reality is a version~I problem.} The perturbation becomes real (modulo gauge) if $\phi_0$ solves
\[
\mathcal P\phi_0=0,\qquad \mathcal K\phi_0=r .
\]
This is the source-free version~I equation of the previous subsection, with $2\bar\psi$ replaced by the residual $r$. It is solvable whenever every homogeneous anti-self-dual field is of the form $\mathcal K\phi_0$, which is the completeness question of Section~\ref{subsec:redundancy}.
\end{enumerate}
In this sense the two versions combine: version~II provides a particular solution carrying the prescribed curvature and the source, and version~I supplies the homogeneous correction needed for reality. The fourth-order operator $\mathcal K$ has therefore not disappeared from version~II. It has been moved from the main construction into a source-free correction. However, the next paragraph argues that for retarded solutions this correction vanishes, $r=0$, so that $\mathcal K$ never has to be inverted.

\paragraph{The role of causality.} In the standard construction, reality comes for free, and causality is used to select the physical solution and to fix the integration constants of the fourth-order inversion. In our framework these roles are distributed differently, and causality can be expected to do more. Consider a real conserved source, and choose retarded solutions at both wave equations: the retarded $\psi$ in step 1 and the retarded $\phi$ in step 2 of version~II. Since $d_3$ and $d_1^*$ are local operators and $j$ is supported on the source, $\delta g_c$ then vanishes to the past of the source. The following argument suggests that $\delta g_c$ is then real modulo gauge.
\begin{enumerate}
\item Since the background, the linearised Einstein operator and the source are real, $\overline{\delta g_c}$ solves the same sourced equation as $\delta g_c$. Hence $\mathrm{Im}\,\delta g_c$ solves the source-free equation.
\item $\mathrm{Im}\,\delta g_c$ also vanishes to the past of the source.
\item Its two curvatures therefore solve homogeneous spin-two equations with vanishing past data, and hence vanish. By the local uniqueness statement above, $\mathrm{Im}\,\delta g_c$ is then pure gauge.
\end{enumerate}
This is the analogue, for conformally flat backgrounds, of the use of Wald's uniqueness theorem in~\cite{Green:2020sourced}; the zero-mode part $\dot g$ is absent here because the background has no parameters, such as mass and angular momentum, that could be perturbed. 

If this argument holds, $\delta g_c$ need not itself be real: it differs from the real perturbation $\mathrm{Re}\,\delta g_c$ by the imaginary pure-gauge term $\mathrm{i}\,\mathrm{Im}\,\delta g_c$. This is, however, all that is needed. Since the background Weyl curvature vanishes, the linearised Weyl curvature is gauge invariant, so a pure-gauge perturbation contributes to neither $\psi$ nor $\tilde\psi$. The curvatures of $\delta g_c$ are therefore exactly those of the real perturbation $\mathrm{Re}\,\delta g_c$, for which the anti-self-dual curvature is the complex conjugate of the self-dual one. The reality condition~\eqref{eq:reality-condition}, which is a statement about curvatures, thus holds exactly, not merely modulo gauge. Conversely, if~\eqref{eq:reality-condition} holds, then $\mathrm{Im}\,\delta g_c$ has vanishing curvatures and, since $\delta g_c$ and $\overline{\delta g_c}$ have the same real source, vanishing source; by the local uniqueness statement above it is pure gauge. Locally on the conformally flat patch, the statements ``$\delta g_c$ is real modulo gauge'' and ``\eqref{eq:reality-condition} holds'' are therefore equivalent. The leftover imaginary part of the metric carries no gauge-invariant information. It can be discarded by passing to $\mathrm{Re}\,\delta g_c$, which has the same curvatures and the same source, or removed by a gauge transformation with complex parameter, which is permitted in the complexified linear theory. In either case the fourth-order equation for $\phi_0$ never needs to be solved. 

The above argument relies on two inputs that we have not established here. The first is the propagation of the first-order Bianchi constraint for the retarded $\psi$, discussed in Section~\ref{sec:sourced-equations}. The second is the uniqueness of retarded solutions modulo gauge on the conformally flat patch; this requires care on the Poincar\'e patch of de Sitter space, where a source may be present for all past times. The example of Section~\ref{sec:example-static-mass} is consistent with this picture. There the retarded curvature satisfies the first-order constraint, and the simplest closed-form solution of $\mathcal P\phi=\psi$ already gives a metric that is real exactly, not only modulo gauge, so that $r=0$ and no fourth-order equation has to be solved. This holds even though, on the Poincar\'e patch, the source is present for all past times. We have not, however, shown that this potential is the retarded solution of $\mathcal P\phi=\psi$, so the example illustrates the argument rather than proving it. For problems with incoming radiation, or with data prescribed on a Cauchy surface, there is no such shortcut, and~\eqref{eq:reality-condition} becomes a condition on the data for $\phi$. These are all subtle questions that we postpone to future work.

\subsection{What is Simplified}\label{subsec:simplification}

The comparison of the three constructions with respect to the three issues (a)--(c) listed at the beginning of this section is summarised in Table~\ref{tab:reconstruction-comparison}.

\begin{table}[ht]
\centering
\small
\renewcommand{\arraystretch}{1.3}
\begin{tabular}{p{0.17\textwidth}p{0.25\textwidth}p{0.22\textwidth}p{0.25\textwidth}}
\hline
 & Standard (Kerr) & Version I & Version II\\
\hline
Potential equation & $\mathcal O_0^*\psi_H=\eta$ & $\mathcal P\phi=\eta$ & $\mathcal P\phi=\psi$\\
(a) Prescribed curvature & fourth-order inversion~\eqref{psi0-psiH} & fourth-order $\mathcal K\phi=2\bar\psi$ & second-order $\mathcal P\phi=\psi$\\
(b) Sources & sourced $\eta$ and corrector $x$, both from integrations along null geodesics; $x$ nonlocal & corrector needed, as in the standard construction & corrector $x=-j$, local and algebraic\\
(c) Reality & by construction (real part) & by construction (real part) & condition~\eqref{eq:reality-condition}; expected automatic for retarded solutions; holds for the static mass of Section~\ref{sec:example-static-mass}\\
\hline
\end{tabular}
\caption{Comparison of the standard Teukolsky--Wald reconstruction on Kerr with the two versions of the first-order reconstruction on constant-curvature backgrounds.}
\label{tab:reconstruction-comparison}
\end{table}

The simplifications produced by the first-order scheme are thus the following. First, and most importantly, the corrector needed to accommodate sources is local and algebraic: it is the term $-j$, whose metric part is proportional to the stress-energy tensor. This is a consequence of the algebraic reverse map from connections to 2-form perturbations, available for $\Lambda\neq0$, together with the choice of the curvature itself as the source of the potential equation. Second, the potential reproducing a prescribed curvature is obtained by solving a second-order equation, rather than by inverting a fourth-order relation. 

The price to pay for these simplifications is as follows. First, version~II needs the full self-dual curvature $\psi$, which must satisfy the first-order Bianchi constraint, before the potential can be found, whereas in the standard construction one works with a single complex scalar. Since a generic curvature, such as the Coulomb field of a point mass (Section~\ref{sec:example-static-mass}), is not aligned with a fixed null direction, the potential then also has all five components, unlike the rank-one potential of version~I. On constant-curvature backgrounds this is a mild cost, since after the conformal rescaling every component obeys a decoupled flat wave equation. Second, reality is not manifest in version~II. It is the fourth-order condition~\eqref{eq:reality-condition}, which we have argued to hold automatically for retarded solutions, and which can otherwise be imposed by adding a solution of version~I, at the cost of a fourth-order inversion similar to that of the standard construction. 

\subsection{Further Remarks: Redundancy and Global Issues}\label{subsec:redundancy}

The potential $\phi^{ij}$ has five complex components, each satisfying a scalar wave equation, while the sector with vanishing chiral curvature carries a single complex radiative degree of freedom. How much of this freedom is needed depends on the problem.

For the source-free radiative perturbations of version~I the construction is highly redundant. Part of this redundancy is gauge: as noted in Sections~\ref{subsec:gauge} and~\ref{sec:conformal-reconstruction}, both $a=\sd\cdot\zeta$ and $a=d^A\omega$, which describe connection perturbations due to a diffeomorphism and a gauge transformation, lie in $\ker d_2$ and produce pure-gauge metric perturbations. The rank-one potentials $\phi^{ij}=\lambda^3S^iS^jf$ of Section~\ref{sec:example-ds} suggest that a single scalar suffices, as for the Hertz potential of the standard construction. We expect this to hold for radiative solutions, with the possible exception of modes propagating along the null direction singled out by $S^i$.

For sourced problems, and for non-radiative fields, aligned potentials do not suffice. By~\eqref{P-flat-Box} a rank-one potential produces only curvature aligned with $S^iS^j$, whereas the curvature~\eqref{eq:sds-psi} of a point mass is of type D and is not aligned with any constant null vector. The potential of version~II then has all five components, see~\eqref{eq:sds-potential}, and is regular away from the worldline. We expect that an aligned description of such fields would require potentials with string-like singularities, as for the radiation-gauge Hertz potentials on Kerr. The non-aligned components thus allow version~II to reproduce such curvatures with a potential that is regular away from the source, consistent with its local corrector.

A characterisation of the redundancy, and a proof that every radiative solution with vanishing chiral curvature arises, locally and modulo gauge, from a rank-one potential, would complete the picture; we leave these questions for future work.

A further remark is that our constructions are local. The conformally flat chart covers only a patch of de Sitter or anti-de Sitter space, for example the Poincar\'e patch, and the weight $\lambda^{-4}$ degenerates at its boundary. The formal adjoints discard boundary terms. For anti-de Sitter space this is precisely where the boundary conditions at the conformal boundary enter, and a global treatment would have to address both points.

Yet another remark is that all algebraic and differential statements of Sections~\ref{sec:linear-eom}--\ref{sec:sourced-equations} hold equally in the real Euclidean setting, with $E=\R^3$ and the factors of $\im$ adjusted to the convention $\star\sd^i=\sd^i$, where the formalism describes $SU(2)$ structures. The round $S^4$ and hyperbolic $H^4$ are conformally flat with $\lambda=(1\pm|x|^2)/2$ in stereographic and Poincar\'e-ball coordinates, respectively. In this setting the sector with vanishing chiral curvature consists of real anti-self-dual Einstein perturbations, and the reality issue discussed above does not arise.

\subsection{Outlook: Backgrounds with Nonvanishing Weyl Curvature}\label{subsec:kerr-outlook}

The algebraic step that recovers the 2-form perturbation from the connection uses only that the curvature equation $F^i=M^{ij}\sd^j$, with $M=\Psi+\frac{\Lambda}{3}\mathbb{I}$, can be solved for the two-forms, $\sd=M^{-1}F$; this is also the condition under which the pure connection formulation exists~\cite{Krasnov:2011pp}, see Appendix~\ref{apx:inner-products}. On constant-curvature backgrounds this is the condition $\Lambda\neq0$. For the Kerr solution, $\Lambda=0$, but the background Weyl curvature is of type D, with $\Psi$ proportional to ${\rm diag}(-1,-1,2)$ in an adapted frame, and $M^{ij}=\Psi^{ij}$ is invertible away from the curvature singularity. On such backgrounds the linearised curvature equation contains the term $\Psi \sigma$, whose five-dimensional component no longer vanishes, so the projections and the principal identity acquire background-curvature corrections. Working these out is the natural next step towards Kerr. 

It would be particularly important to know whether the local corrector of version~II survives on Kerr. Since $M^{ij}=\Psi^{ij}$ is invertible there, the reverse map from connections to 2-form perturbations remains algebraic, which suggests that it might; the background-curvature corrections to the principal identity, and the coupling of the two helicities by the background Weyl curvature, which is absent on the constant-curvature backgrounds considered here (Section~\ref{subsec:version-I}), would have to be taken into account. A further issue is that version~II uses the full curvature $\psi$, whereas on Kerr only its extreme components $\psi_0$ and $\psi_4$ obey decoupled equations; the version~II scheme therefore cannot simply be transplanted to Kerr, and the role of the remaining components must be understood. We leave this to future work.

\section*{Acknowledgements} Both authors are supported by the Science and Technology Facilities Council, K.K. through grant ST/C004235/1 and A.S. through grant ST/X000672/1. The authors are grateful to Stephen Green and Jacob Neef for a helpful discussion. The AI system Claude Opus~5.5 (Anthropic), used through the Claude Science research environment, assisted in the preparation of this paper. It helped to organise the exposition, starting from notes written by the authors; verified many of the formulas by independent computer-algebra calculations; carried out the computations for the static mass in de Sitter space of Section~\ref{sec:example-static-mass}; and drafted the comparison with the standard Teukolsky--Wald reconstruction, including the discussion of reality, in Section~\ref{sec:discussion}. All AI-generated material, including the formulas, has been checked by the authors, who take full responsibility for the content of this paper.

\appendix

\section{Metric reality conditions in the chiral formalism}\label{apx:metric-reality}

A real Lorentzian metric is represented in the chiral formalism by a complex triple of self-dual two-forms. Accordingly, taking the componentwise real part of a two-form perturbation is not the operation that takes the real part of its metric. In this appendix we describe the appropriate algebraic operation. It requires no differentiation or inversion of a field equation and applies, in particular, to the single-helicity perturbations obtained from Hertz potentials.

Fix a real Lorentzian background metric $g$ and a normalised self-dual triple $\Sigma^i$, with $\bar\Sigma^i=\overline{\Sigma^i}$ its conjugate anti-self-dual triple. All contractions below use the background metric. A perturbation satisfying linearised metricity, $P_5\sigma=0$, has the decomposition
\begin{equation}\label{eq:reality-decomposition}
 \sigma^i=s\Sigma^i+\epsilon^{ijk}u^k\Sigma^j
                  +\theta^{ij}\bar\Sigma^j,
 \qquad
 s=\frac1{12}\Sigma^{i\mu\nu}\sigma^i_{\mu\nu},
 \qquad
 \theta^{ij}=\frac14\bar\Sigma^{j\mu\nu}\sigma^i_{\mu\nu}.
\end{equation}
The scalar $s$ and the general complex matrix $\theta$ encode the metric perturbation; $u^i$ describes an infinitesimal internal frame rotation and does not affect the metric.

\paragraph{Linearised reality conditions.}
Linearising the orthogonality condition for the perturbed triple and its complex conjugate gives
\begin{equation}\label{eq:linearised-reality-orthogonality}
 \sigma^i\wedge\bar\Sigma^j
       +\Sigma^i\wedge\overline{\sigma^j}=0.
\end{equation}
Write $\Sigma^i\wedge\Sigma^j=\delta^{ij}W$. For the real Lorentzian background, $W$ is purely imaginary, so that
$\bar\Sigma^i\wedge\bar\Sigma^j=-\delta^{ij}W$.
Substituting \eqref{eq:reality-decomposition} into
\eqref{eq:linearised-reality-orthogonality} therefore gives
\begin{equation}
 \bigl(-\theta^{ij}+\overline{\theta^{ji}}\bigr)W=0.
\end{equation}
Thus the conformal metric reality condition is that $\theta$ be Hermitian. The volume reality condition additionally requires $s$ to be real: indeed,
$\delta(\Sigma^i\wedge\Sigma^i)=6sW$.
The complete metric reality conditions at first order are consequently
\begin{equation}\label{eq:linearised-metric-reality}
 s=\bar s,\qquad \theta=\theta^\dagger,
 \qquad (\theta^\dagger)^{ij}:=\overline{\theta^{ji}}.
\end{equation}
There is no restriction on the internal rotation $u^i$. 

\paragraph{An algebraic real-metric projection.}
For an arbitrary complex perturbation satisfying linearised metricity, define
\begin{equation}\label{eq:two-form-real-projection}
 (\mathcal R\sigma)^i
   :=\operatorname{Re}(s)\,\Sigma^i
       +\frac12\bigl(\theta^{ij}+\overline{\theta^{ji}}\bigr)
           \bar\Sigma^j .
\end{equation}
This chooses the representative with vanishing internal rotation. An arbitrary internal rotation may be added without changing the resulting metric. The map $\mathcal R$ is real-linear and idempotent; on the quotient by internal rotations it is the projection onto perturbations with real metric.

To verify that this operation gives precisely the real part of the original metric, introduce the symmetric background tensors
\begin{equation}
 K^{ij}_{\mu\nu}
   :=\Sigma^i_{(\mu}{}^\rho\bar\Sigma^j_{\nu)\rho}.
\end{equation}
They satisfy $\overline{K^{ij}_{\mu\nu}}=K^{ji}_{\mu\nu}$. Applying the metric-extraction map gives
\begin{equation}
 P_{10}(\sigma)_{\mu\nu}
   =s\,g_{\mu\nu}+\theta^{ij}K^{ij}_{\mu\nu},
\end{equation}
where the internal rotation drops out. Complex conjugation therefore acts on the metric coefficients by $s\mapsto\bar s$ and $\theta\mapsto\theta^\dagger$, and hence
\begin{equation}\label{eq:real-projection-metric}
 P_{10}(\mathcal R\sigma)
     =\operatorname{Re}\bigl(P_{10}(\sigma)\bigr).
\end{equation}
The operation \eqref{eq:two-form-real-projection} thus implements taking the real part of the metric directly on the two-form variable. The resulting two-forms remain complex, as required in the Lorentzian chiral formalism. In particular, $\mathcal R\sigma$ should not be confused with
$\tfrac12(\sigma+\overline{\sigma})$.

\section{Decomposition of Differential Forms}\label{sec:decomp-of-diff-forms}

The first-order maps $d_1, d_2$ introduced in Section~\ref{sec:linear-eom} act on $E$-valued one-forms and two-forms. Before differentiating these fields, it is useful to separate their algebraic components. All decompositions in this section are pointwise: they depend on the background triple $\sd$, and involve no equations of motion. We use the complexified spaces appropriate to the Lorentzian chiral formulation; the same algebra applies to $\R^3$-valued forms in the real Euclidean setting~\cite{Bhoja:2024xbe}.

The subscripts on the summands below denote their dimensions. The essential distinction for the subsequent analysis is between the five-dimensional curvature component and its thirteen-dimensional complement, which contains the admissible 2-form perturbations. The one-form decomposition will also identify the part of the connection produced by a tracefree potential.

\subsection{\texorpdfstring{$\Eoneforms$}{1-Form With Values} Irreducible Decomposition}

An $E$-valued one-form has twelve components. The background triple defines the algebraic operator $J_1$ 
\begin{align}
	J_1(\theta)^i_\mu = \epsilon^{ijk} \sd^j_\mu{}^\nu \theta^k_\nu.
\end{align}
Using the quaternion algebra \cref{eq:quat-alg} we find that $J_1$ satisfies the following polynomial,
\begin{align}
	(J_1 + \mathbb{I})(J_1 - 2\mathbb{I}) = 0.
\end{align}
The resulting decomposition is
\begin{align}
	\Eoneforms = (\Eoneforms){}_4 \oplus (\Eoneforms){}_8.
\end{align}

The summands have dimensions four and eight, with eigenvalues $+2$ and $-1$, respectively. The four-dimensional part is isomorphic to the space of ordinary one-forms. Contraction with the background triple extracts the corresponding one-form, up to normalisation:
\begin{align}
	\sd^i_\mu{}^\nu \theta^i_\nu \in \Lambda^1 \cong (\Eoneforms){}_4.
\end{align}
Conversely, an ordinary one-form determines an element of the four-dimensional summand by
\begin{align}
	\xi_\mu \in \Lambda^1, \quad (\sd^i \cdot \xi){}_\mu = \sd^i_\mu{}^\nu \xi_\nu \in (\Eoneforms){}_4 \subset \Eoneforms. 
\end{align}
The quaternion algebra verifies that this element lies in the claimed eigenspace:
\begin{align}
	J_1( \sd \cdot \xi){}^i = 2 \sd^i \cdot \xi.
\end{align}

\subsection{\texorpdfstring{$\Etwoforms$}{2-Forms With Values} Irreducible Decomposition}\label{subsec:2-form-irrep-decomp}

For two-forms, we first split them into self-dual and anti-self-dual parts. The self-dual part can be written as a matrix multiplying $\sd^j$; this matrix splits into its trace, antisymmetric, and symmetric tracefree parts. The anti-self-dual part is a general matrix multiplying $\asd^j$. Together these give dimensions $1+3+5+9=18$. The decomposition can also be organised using the algebraic operator
\begin{align}
	J_2(B){}^i_{\mu\nu} = \epsilon^{ijk} \sd^j_{[\mu|}{}^\rho B^k_{|\rho|\nu]},
\end{align}

for which the decomposition becomes,
\begin{align}
	E \otimes \Lambda^2 = (\Etwoforms){}_1 \oplus (\Etwoforms){}_3 \oplus (\Etwoforms){}_5 \oplus (\Etwoforms){}_9.
\end{align}

Representatives of the four summands are
\begin{align}
	& h \in \C, \quad h \sd^i \in (\Etwoforms){}_1, \\
	& h^i \in \C^3, \quad \epsilon^{ijk} h^k \sd^j \in (\Etwoforms){}_3, \\
	& h^{ij} \in {\rm Sym}^2_0(\C^3), \quad h^{ij} \sd^j \in (\Etwoforms){}_5, \\
	& \theta^{ij} \in {\rm Mat}(3,\C), \quad \theta^{ij} \asd^j \in (\Etwoforms){}_9 \cong E \otimes \Lambda^-.
\end{align}

Here ${\rm Mat}(3,\C)$ is the space of complex $3\times3$ matrices and ${\rm Sym}^2_0(\C^3)$ its symmetric tracefree subspace. The one-dimensional and nine-dimensional pieces together encode the metric variation, the three-dimensional piece describes internal frame rotations, and the five-dimensional piece is excluded by linearised metricity. Angle brackets on internal indices denote symmetric tracefree projection; indices between vertical bars are omitted from that projection. The components of an arbitrary $B\in E\otimes\Lambda^2$ are extracted by the following contractions:
\begin{align}
	h & = \frac{1}{12}\sd^{i\mu\nu} B^i_{\mu\nu}, \\
	h^i & = -\frac{1}{8} \epsilon^{ijk} \sd^{j\mu\nu} B^k_{\mu\nu}, \\
	h^{ij} & = \frac{1}{4} \sd^{\langle i | \mu\nu} B^{| j \rangle}_{\mu\nu}, \label{eq:sym-0-C3-extract}\\
	\theta^{ij} & = \frac{1}{4} \asd^{j\mu\nu} B^i_{\mu\nu}. 
\end{align}
Reinserting these coefficient fields into their respective basis two-forms gives the algebraic projections. In particular, $P_5$ extracts the symmetric tracefree self-dual component. A linearised 2-form perturbation obeys $P_5\sigma=0$, whereas the chiral Weyl perturbation lies in the image of $P_5$.

\section{Computation of the Adjoints}\label{apx:adjoints}

In this appendix we verify the adjoint relations stated in Section~\ref{sec:linear-eom}, using the pairings defined there, and derive the simplified form \eqref{d2-star} of $d_2^*$.

Integration by parts gives
\begin{align}
	\langle a, d_1(\sigma) \rangle = \im (a^i, \im \star d^A \sigma^i) = (d^A a^i, \star \sigma^i).
\end{align}
Because the 2-form perturbation has no five-dimensional component, only the complementary projection contributes to this pairing:
\begin{align}
	\langle a, d_1(\sigma) \rangle = ((\mathbb{I} - P_5)(d^A a){}^i, \star \sigma^i) = \langle \frac{3}{\Lambda}(\mathbb{I} - P_5)(d^A a), \sigma \rangle.
\end{align}
This verifies the adjoint relation for $d_1$. 

For the curvature pairing, pairing a curvature-type field with $d_2a$ gives
\begin{align}
	\langle \psi, d_2(a) \rangle = -\frac{3}{\Lambda} (\psi^{ij} \sd^j, \star P_5(d^A a){}^i ) = -\frac{3\im}{\Lambda} (\psi^{ij} \sd^j, P_5(d^A a){}^i).
\end{align}
The algebraic projection $P_5$ is self-adjoint for the metric pairing. Indeed, using $P_5(B)^i_{\mu\nu}=\frac14 B^{\langle i}_{\rho\sigma}\sd^{j\rangle\rho\sigma}\sd^j_{\mu\nu}$, one checks
\begin{align}
	(B_1^i, P_5(B_2){}^i) = \frac12 B_1^{i\mu\nu} \frac{1}{4} B^{\langle i}_{2\,\rho\sigma} \sd^{j\rangle\rho\sigma} \sd^j_{\mu\nu} = \frac{1}{8} B^{\langle i}_{1\,\rho\sigma} \sd^{j\rangle\rho\sigma} \sd^{j\mu\nu} B^i_{2\,\mu\nu} = (P_5(B_1){}^i, B_2^i).
\end{align}
Therefore, $P_5^* = P_5$ and the adjoint of $d_2$ becomes
\begin{align}
	\langle \psi, d_2(a) \rangle & = -\frac{3\im}{\Lambda} (\psi^{ij} \sd^j, d^A a^i) = -\frac{3\im}{\Lambda} (\star d^A \star (\psi^{ij} \sd^j), a^i) 
	= \frac{3}{\Lambda}(\star d^A (\psi^{ij} \sd^j), a^i) \\
	& = \frac{3 \im}{\Lambda} (-\im (J_1 J_1^{-1} \star d^A)(\psi \sd){}^i, a^i) = \langle -\frac{3\im}{\Lambda} (J_1^{-1} \star d^A)(\psi \sd){}^i, a^i \rangle
\end{align}
Here we used $(B^i,d^Aa^i)=(\star d^A\star B^i,a^i)$, modulo boundary terms, for $E$-valued two-forms $B$; this is the same integration by parts as in the computation of the adjoint of $d_1$ above. Hence,
\begin{align}
	d_2^*(\psi){}^i = -\frac{3\im}{\Lambda}J_1^{-1}(\star d^A (\psi \sd)){}^i.
\end{align}

To simplify this expression, we use \eqref{a-8}. This is checked by an explicit calculation, using the definition of $J_1$ and the quaternion algebra of $\Sigma$'s. Indeed, writing $X^i=\sd^j\cdot v^{ij}$ with $v^{ij}$ a symmetric tracefree matrix of one-forms, one finds
\[
J_1(X)^i=\epsilon^{ijk}\sd^j\cdot\sd^l\cdot v^{kl}
=\epsilon^{ijk}\left(-\delta^{jl}+\epsilon^{jlm}\sd^m\cdot\right)v^{kl}
=-\epsilon^{ijk}v^{kj}+\sd^i\cdot v^{kk}-\sd^k\cdot v^{ki}=-X^i .
\]

We then have that
\begin{align}
	\star d^A (\psi^{ij} \sd^j){}_\mu = \frac{3}{3!} \epsilon^{\nu\rho\sigma}{}_\mu \nabla^A_\nu \psi^{ij} \sd^j_{\rho\sigma} = -\im \sd^j_\mu{}^\nu \nabla^A_\nu \psi^{ij} = -\im \sd^j \cdot d^A \psi^{ij}.
\end{align}
Here $\nabla^A_\mu X^i_\nu=\nabla_\mu X^i_\nu+\epsilon^{ijk}A^j_\mu X^k_\nu$ includes both the Levi-Civita connection and the internal connection; it acts on both internal indices of $\psi^{ij}$. We also used the fact that $\nabla^A_\mu \Sigma^i_{\rho\sigma}=0$, as well as the self-duality of $\Sigma$. Since $J_1$ acts as $-1$ on the eight-dimensional summand, its inverse can be eliminated:
\begin{align}\label{eq:J-1-inv-star-d-A-id}
	(J_1^{-1} \star d^A)(\psi \sd){}^i = \im \sd^j \cdot d^A \psi^{ij}.
\end{align}
Substituting this into the expression for $d_2^*$ gives \eqref{d2-star}.

\section{Derivation of Connection Equations}\label{apx:torsion-free-sol}

Here we explicitly describe the process of solving for $A^i$ in \cref{eq:torsion-free}. Taking the Hodge star of the second term in \cref{eq:torsion-free} gives
\begin{align}
	\star (\epsilon^{ijk} A^j \wedge \sd^k){}_\mu & = \frac{1}{3!} \epsilon^{\nu\rho\sigma}{}_\mu \epsilon^{ijk} (A^j \wedge \sd^k){}_{\nu\rho\sigma} =  \frac{1}{2!} \epsilon^{\nu\rho\sigma}{}_\mu \epsilon^{ijk} A^j_\nu \sd^k_{\rho\sigma} \\
	& =  \frac{1}{2} \epsilon^{ijk}\epsilon_{\mu}{}^{\nu\rho\sigma} \sd^j_{\rho\sigma} A^k_\nu = \im \epsilon^{ijk} \sd^j_\mu{}^\nu A^k_\nu = \im J_1{(A)}^i_\mu.
\end{align}
Using the quaternion algebra satisfied by $\sd^i$, see~\cite{Bhoja:2024xbe}, we find that $J_1^{-1} = \frac{1}{2} (J_1 - \mathbb{I} )$. Combining this with the first term in \cref{eq:torsion-free} we have,
\begin{align}
	\star d \sd^i + \im J_1{(A)}^i = 0. \quad \Rightarrow \quad A^i = \im J_1^{-1}{(\star d \sd)}^i.
\end{align}

\section{Inner Products from the Action}\label{apx:inner-products}

We fix the relative signs and normalisations of the quadratic terms before identifying the bilinear pairings used for formal adjoints. All equalities between actions that involve integration by parts are understood modulo boundary terms.

With the curvature convention of Section~\ref{sec:pelb-form-of-gr}, the chiral Lagrangian is
\begin{align}
 L=\sd^i\wedge F^i-\frac12
       \left(\Psi^{ij}+\frac\Lambda3\delta^{ij}\right)\sd^i\wedge\sd^j.
\end{align}
Variation with respect to $\sd$ gives $F^i=(\Psi^{ij}+\frac\Lambda3\delta^{ij})\sd^j$, with the required sign. Let $L^{(2)}$ denote the coefficient of the quadratic perturbation, rather than twice that coefficient. Expanding about $\Psi=0$ gives
\begin{align}
 L^{(2)}={}&\sigma^i\wedge d^Aa^i
       +\frac12\epsilon^{ijk}\sd^i\wedge a^j\wedge a^k
       -\frac\Lambda6\sigma^i\wedge\sigma^i
       -\psi^{ij}\sd^j\wedge\sigma^i.
\end{align}
The algebraic quadratic terms determine the pairings up to the sign conventions used to write the field equations. We use
\begin{align}
 \langle a,a\rangle
   &=-\star(\epsilon^{ijk}\sd^i\wedge a^j\wedge a^k)
     =\im(a^i,J_1a^i),\\
 \langle\sigma,\sigma\rangle
   &=\frac\Lambda3\star(\sigma^i\wedge\sigma^i)
     =\frac\Lambda3(\sigma^i,\star\sigma^i).
\end{align}
These are precisely the pairings for which the reverse map to 2-forms is the formal adjoint of the compatible-connection map. The factors of $\im$ follow from the Lorentzian self-duality convention $\star\sd^i=\im\sd^i$.

To obtain the curvature pairing, eliminate the two-forms. Write $M=\Psi+\frac\Lambda3\mathbb I$. Wherever $M$ is invertible,
\begin{align}\label{eq:pure-con-sd-sol}
 F^i=M^{ij}\sd^j,\qquad \sd^i=(M^{-1})^{ij}F^j.
\end{align}
Substitution gives the first-order pure connection Lagrangian
\begin{align}
 L_{\rm pc}=\frac12(M^{-1})^{ij}F^i\wedge F^j.
\end{align}
In the following expansion $M$ and $F$ denote the background values and $\delta M=\psi$. Matrix multiplication and the expansion of $F(A+a)$ give
\begin{align}
 L_{\rm pc}^{(2)}={}&\frac12(M^{-1})^{ij}d^Aa^i\wedge d^Aa^j
 +\frac12(M^{-1})^{ij}F^i\wedge\epsilon^{jkl}a^k\wedge a^l\nonumber\\
 &-(M^{-1}\psi M^{-1})^{ij}F^i\wedge d^Aa^j
 +\frac12(M^{-1}\psi M^{-1}\psi M^{-1})^{ij}F^i\wedge F^j.
\end{align}
Using $M=\frac\Lambda3\mathbb I$ and $F^i=\frac\Lambda3\sd^i$, this becomes
\begin{align}
 L_{\rm pc}^{(2)}={}&\frac12\epsilon^{ijk}\sd^i\wedge a^j\wedge a^k
       +\frac3{2\Lambda}d^Aa^i\wedge d^Aa^i\nonumber\\
 &-\frac3\Lambda\psi^{ij}\sd^j\wedge d^Aa^i
       +\frac3{2\Lambda}\psi^{ij}\psi^{ik}\sd^j\wedge\sd^k.
\end{align}
The first two terms cancel modulo an exact form, since
\begin{align}
 d^Aa^i\wedge d^Aa^i
 &=d(a^i\wedge d^Aa^i)+a^i\wedge(d^A)^2a^i\nonumber\\
 &=d(a^i\wedge d^Aa^i)-\frac\Lambda3\epsilon^{ijk}\sd^i\wedge a^j\wedge a^k.
\end{align}
Thus
\begin{align}
 L_{\rm pc}^{(2)}\equiv
 -\frac3\Lambda\psi^{ij}\sd^j\wedge d^Aa^i
 +\frac3{2\Lambda}\psi^{ij}\psi^{ik}\sd^j\wedge\sd^k.
\end{align}
Following the same rule as for the connection and 2-form pairings above, where each pairing is $-2$ times the coefficient of the corresponding quadratic term in the Lagrangian, the quadratic curvature term defines the pairing
\begin{align}
 \langle\psi,\psi\rangle
 &=-\frac3\Lambda\star(\psi^{ij}\sd^j\wedge\psi^{ik}\sd^k)
 =-\frac3\Lambda(\psi^{ij}\sd^j,\star(\psi^{ik}\sd^k))\nonumber\\
 &=-\frac{3\im}\Lambda(\psi^{ij}\sd^j,\psi^{ik}\sd^k).
\end{align}

\section{Linearised Equation of Motion for \texorpdfstring{$\Lambda = 0$}{Λ = 0}}\label{sec:lin-eom-lambda-0}

This appendix treats the flat background, $\Lambda=0$. Its message differs from that of the main text: there is no equivalent connection description, and the first-order maps generate only a single-helicity sector.

Setting $\Lambda=0$ on a constant-curvature background gives a flat background. The first-order compatibility equation remains unchanged, but the algebraic step that recovered $\sigma$ by dividing by $\Lambda$ is no longer available. The curvature equation becomes
\begin{align}
	d^A a^i = \psi^{ij} \sd^j.
\end{align}
The right-hand side lies entirely in the five-dimensional Weyl summand. Its two projections give
\begin{align}
	\psi^{ij} \sd^j = P_5(d^A a){}^i, \\
	(\mathbb{I} - P_5)(d^A a){}^i = 0.
\end{align}
The first equation still extracts the chiral curvature. The second is now a first-order constraint on the physical connection perturbation; it no longer determines the 2-form perturbation. Introduce the unnormalised operator
\begin{gather}
	\psi^{ij} \sd^j = d_2(a){}^i, \\
	\tilde{d}_1^* (a){}^i = 0, \quad \tilde{d}_1^* = (\mathbb{I}-P_5) d^A.
\end{gather}
The tilde distinguishes these operators from the $\Lambda$-dependent adjoints of the preceding sections.

\paragraph{Pairings.} At $\Lambda=0$ the algebraic $\sigma\sigma$ term in the action vanishes, and the elimination of the two-forms using $M^{-1}$ of Appendix~\ref{apx:inner-products} is unavailable on the background. The connection pairing $\langle a,a\rangle=\im(a^i,J_1a^i)$ is still defined by its quadratic action term, but the 2-form and curvature pairings must be chosen independently. Removing the $\Lambda$-dependent prefactors from the pairings of Appendix~\ref{apx:inner-products}, but retaining their signs, we use
\begin{align}\label{flat-pairings}
 \langle\sigma,\sigma\rangle_0=(\sigma^i,\star\sigma^i),\qquad
 \langle\psi,\psi\rangle_0=-(\psi^{ij}\sd^j,\star(\psi^{ik}\sd^k)).
\end{align}
Formal adjoints are again understood after integration by parts, and the corresponding bilinear forms are obtained by polarisation.

Since the other equations remain unchanged we still have the same operator for $d_1$, that is
\begin{align}
	a^i = d_1(\sigma){}^i = \im J_1^{-1}(\star d^A \sigma){}^i.
\end{align}
Using the previous calculations it is possible to check that $\tilde{d}_1^*$ and $d_1$ are indeed adjoints with respect to the inner product (\ref{flat-pairings}). The adjoint for $d_2$ with respect to this inner product is
\begin{align}
	\tilde{d}_2^*(\psi){}^i = -\im J_1^{-1} \star d^A (\psi \sd){}^i.
\end{align}

\subsection{2-Form Equations and an Auxiliary Connection Construction}

Eliminating the compatible connection from the linearised equations gives the 2-form Einstein operator
\begin{align}
	E_\sigma (\sigma) = (\tilde{d}_1^* d_1)(\sigma){}^i = 0.
\end{align}
For $\Lambda\neq0$, the reverse map followed from the field equations and inverted $d_1$ on solutions. That argument is unavailable here. Instead, use the following expression as a solution-generating ansatz:
\begin{align}
	\sigma^i := \tilde{d}_1^*(a)^i,
\end{align}
and impose the auxiliary second-order equation
\begin{align}
	E_a(a){}^i = (d_1 \tilde{d}_1^*)(a)^i = 0.
\end{align}
Using our metric reconstruction operator we find that,
\begin{align}
	(E_\sigma \tilde{d}_1^*)(a) = (\tilde{d}_1^* E_a)(a).
\end{align}
This is simply associativity of the composition $\tilde d_1^*d_1\tilde d_1^*$. Thus a solution of the auxiliary connection equation generates a 2-form solution. This implication is weaker than the equivalence at nonzero $\Lambda$: $d_1\sigma=d_1\tilde d_1^*a=0$ for the generated 2-form perturbation, so the auxiliary $a$ is not in general its compatible connection. 

The metric perturbations generated in this way correspond to a single graviton helicity: their linearised self-dual Weyl curvature vanishes. In Euclidean signature they deform the flat background to a hyper-K\"ahler one; in the complexified Lorentzian setting they are the complex anti-self-dual, or heavenly, perturbations. Indeed, the hyper-K\"ahler metrics are described in our formalism by triples of 2-forms $\Sigma^i$ that are closed $d\Sigma^i=0$ as well as satisfy $\Sigma^i\wedge \Sigma^j \sim \delta^{ij}$. For such 2-forms the chiral connection $A^i$ vanishes. At the linearised level, such hyper-K\"ahler perturbations are those with $d\sigma^i=0$, and in particular $d_1 \sigma=0$. Perturbations of the form $\sigma=\tilde d_1^*a$ with $E_aa=0$ are of this type, since $d_1\sigma=E_aa=0$. We expect that, conversely, every such perturbation arises in this way locally, by a Poincar\'e-lemma argument using the gauge freedom of $a$, but we do not prove this here.

\subsection{Curvature-Type Potentials}

A further potential construction is obtained by applying the curvature adjoint. In the following formulas $\psi$ denotes a potential with curvature-type indices:
\begin{align}
	a^i = \tilde{d}_2^*(\psi)^i,
\end{align}
The flat-background counterpart of the principal identity is
\begin{align}
	\tilde{d}_2^* d_2 &= -\im J_1^{-1} \star d^A P_5 d^A, \\
	d_1 \tilde{d}_1^* &= \im J_1^{-1} \star d^A (\mathbb{I} - P_5) d^A = -\im J_1^{-1} \star d^A P_5 d^A.
\end{align}
Since the background chiral connection is flat, $(d^A)^2=0$. The two compositions are therefore equal, $\tilde d_2^*d_2=d_1\tilde d_1^*$, without an identity term. We use this equality to verify the ansatz
\begin{align}
	a^i := \tilde{d}_2^*(\psi)^i.
\end{align}
with potential equation
\begin{align}
	E_\psi (\psi){}^{ij} = (d_2 \tilde{d}_2^*)(\psi){}^{ij} = 0.
\end{align}

Given a $\psi^{ij}$ that is a solution to $E_\psi = 0$ we find the reconstructed connection satisfies
\begin{align}
	(E_a \tilde{d}_2^*)(\psi){}^i = (d_1 \tilde{d}_1^* \tilde{d}_2^*)(\psi){}^i = (\tilde{d}_2^* d_2 \tilde{d}_2^*)(\psi){}^i = (\tilde{d}_2^* E_\psi)(\psi){}^i = 0.
\end{align}
The potential thus generates an auxiliary connection field and then a 2-form solution by the preceding map. As in the nonzero-$\Lambda$ construction, it should not be identified with the extracted curvature of the generated field. The second-order potential operator simplifies to
\begin{align}
	(d_2 \tilde{d}_2^*)(\psi){}^{ij} = -\frac{1}{2} \Box^A \psi^{ij}.
\end{align}
On a simply connected patch, flatness of $A$ permits a gauge with $A=0$. Since the background metric in this section is also flat, the potential equation is then the ordinary componentwise wave equation. No conformal rescaling is needed.

\section{Weyl Curvature of the Hertz Metric}\label{apx:hertz-alignment}

This appendix gives the flat-space argument behind the identity $\psi_0[h^c]=0$ used in Section~\ref{subsec:reality-comparison}. Let $o_A$ be the spinor with $l^\mu=o^A\bar o^{A'}$; in flat space it may be taken constant. Up to normalisation, the ingoing radiation gauge Hertz potential corresponds to the totally symmetric spinor potential
\[
\chi_{ABCD}=o_Ao_Bo_Co_D\,\psi_H ,
\]
which is aligned with $o_A$, and the complex Hertz metric is obtained from it by two derivatives, contracted on the unprimed indices,
\[
h^c_{AA'BB'}\propto\nabla^C{}_{(A'}\nabla^D{}_{B')}\,\chi_{ABCD} .
\]
Since $o^A\chi_{ABCD}=0$ and $\chi_{ABCD}$ is symmetric, this metric satisfies the radiation gauge conditions $l^\mu h^c_{\mu\nu}=0$ and $g^{\mu\nu}h^c_{\mu\nu}=0$. The self-dual Weyl spinor of a metric perturbation is $\psi_{ABCD}[h]\propto\nabla_{(A}{}^{A'}\nabla_B{}^{B'}h_{CD)A'B'}$. Substituting $h^c$, each primed index now contracts a pair of derivatives, and in flat space $\nabla_A{}^{A'}\nabla^E{}_{A'}\propto\delta_A{}^E\,\Box$. Hence
\[
\psi_{ABCD}[h^c]\propto\Box^2\chi_{ABCD}=o_Ao_Bo_Co_D\,\Box^2\psi_H .
\]
Contracting with $o^Ao^Bo^Co^D$ gives $\psi_0[h^c]=0$, because $o_Ao^A=0$, and the same argument gives $\psi_1[h^c]=\psi_2[h^c]=\psi_3[h^c]=0$. These identities are algebraic consequences of the alignment of the potential with $o_A$: they hold for any function $\psi_H$, without using its field equation. The only self-dual component that can survive is $\psi_4[h^c]\propto\Box^2\psi_H$, and it vanishes once $\psi_H$ satisfies its field equation, which in flat space is the wave equation. The anti-self-dual curvature, in contrast,
\[
\tilde\psi_{A'B'C'D'}[h^c]\propto\nabla^A{}_{A'}\nabla^B{}_{B'}\nabla^C{}_{C'}\nabla^D{}_{D'}\,\chi_{ABCD} ,
\]
is generically nonzero; its component along $\bar o^{A'}\bar o^{B'}\bar o^{C'}\bar o^{D'}$ is proportional to $l^4\psi_H$. More generally, each derivative is contracted with $o^A$, so contracting a primed index with $\bar o^{A'}$ gives the derivative along $l=o^A\bar o^{A'}$, and contracting it with $\bar\iota^{A'}$ gives the derivative along the complex null vector $o^A\bar\iota^{A'}$. The component $\tilde\psi_k[h^c]$ with $k$ primed indices contracted with $\bar\iota^{A'}$ is therefore proportional to $(4-k)$ derivatives of $\psi_H$ along $l$ and $k$ along $o^A\bar\iota^{A'}$. None of these vanishes in general. On Kerr the same structure persists, with $o_A$ the principal spinor: $\psi_0[h^c]=0$ still follows from the alignment of the potential.

\printbibliography
	
\end{document}